\documentclass{article}

\usepackage{jmlr}

\usepackage[utf8]{inputenc} % allow utf-8 input
\usepackage[T1]{fontenc}    % use 8-bit T1 fonts
\usepackage{hyperref}       % hyperlinks
\usepackage{url}            % simple URL typesetting
\usepackage{booktabs}       % professional-quality tables
\usepackage{amsmath}
\usepackage{amsfonts}       % blackboard math symbols
\usepackage{nicefrac}       % compact symbols for 1/2, etc.
\usepackage{microtype}      % microtypography
\usepackage{lipsum}

\begin{document}

%\title{Doubly robust, machine learning effect estimation in real-world clinical sciences: A practical evaluation of performance in molecular epidemiology cohort settings}

\title{REFINE2: A tool to evaluate real-world performance of machine-learning based effect estimators for molecular and clinical studies}

% \author{
%   Xiang Meng\thanks{Use footnote for providing further
%     information about author (webpage, alternative
%     address)---\emph{not} for acknowledging funding agencies.} \\
%   Department of Statistics\\
%   Harvard University\\
%   Cambridge, MA 02138 \\
%   \texttt{xmeng@g.harvard.edu} \\
%   %% examples of more authors
%   \And
%  Jonathan Huang \\
%   TBD\\
%   \texttt{Jonathan\_Huang@sics.a-star.edu.sg} \\
% }

\author{\name Xiang Meng \email xmeng@g.harvard.edu \\
       \addr Department of Statistics\\
  Harvard University\\
  Cambridge, MA 02138, USA \\
       \AND
       \name Jonathan Huang \email Jonathan\_Huang@sics.a-star.edu.sg \\
       \addr Biostatistics and Human Development \\
       Singapore Institute for Clinical Sciences  \\
       Agency for Science, Technology and Research \\
       Singapore 117609, Singapore \\
       \\
       Centre for Quantitative Medicine \\
       Duke-NUS Medical School \\
       Singapore 169857, Singapore}

\editor{-}

\maketitle

% Most recent version: 2021 May 05
% Jon edited 2021 Aug 24
%%% new comments are tagged with: NEW_AUG
% Xiang edited 2021 Sep 21
%%% new comments are tagged with: XM_AUG

%%% XM_AUG: Shortened the abstract. Move some sentences from abstract to Introduction
%%% JYH: New abstract for a Nature Methods submission
\begin{abstract}
Data-adaptive (machine learning-based) effect estimators are increasingly popular to reduce bias in high-dimensional bioinformatic and clinical studies (e.g. real-world data, target trials, -omic discovery). Their relative statistical efficiency (high power) is particularly invaluable in these contexts since sample sizes are often limited due to practical and cost concerns. However, these methods are subject to technical limitations that are dataset specific and involve computational trade-offs. Thus, it is challenging for analysts to identify when such methods may offer benefits or select amongst statistical methods. We present extensive simulation studies of several cutting-edge estimators, evaluating both performance and computation time. Critically, rather than use arbitrary simulation data, we generate synthetic datasets mimicking the observed data structure (plasmode simulation) of a real molecular epidemiologic cohort. We find that machine learning approaches may not always be indicated in such data settings, but that performance is highly context dependent. We present a user-friendly Shiny app REFINE2 (Realistic Evaluations of Finite sample INference using Efficient Estimators) that enables analysts to simulate synthetic data from their own datasets and directly evaluate the performance of several cutting-edge algorithms in those settings. This tool may greatly facilitate the proper selection and implementation of machine-learning-based effect estimators in bioinformatic and clinical study contexts.
\end{abstract}


% JYH: Revised for Nat Methods 11/FEB/2022
\section{Introduction}
The proper application of machine learning to infer biological effects from multiomic data presents an on-going challenge in bioinformatic and clinical studies (\cite{Libbrecht2015-zn}, \cite{Hill2016-qa}, \cite{LI2010493}, \cite{10.1145/3307339.3343177}, \cite{10.1093/bib/bbk007}, \cite{10.3389/fbinf.2021.746712},\cite{Prosperi2020},  \cite{Park2021-gy}). Estimation of treatment effects via data-adaptive, efficient estimators such as Targeted Minimum Loss Estimation (\cite{Diaz2017-re}, \cite{Diaz2020-sh}) and related approaches to model and compute counterfactual quantities based on Bayesian networks have risen as state-of-the-art in both genomic (\cite{Gruber2010-je}, \cite{Ness2016-qr}, \cite{White2019-qh}) and medical applications (\cite{Balzer2016-bh}, \cite{Kreif2017-lq}, \cite{Sofrygin2019-dl}, \cite{Rossides2021-ns}, \cite{Huang2021-mn}). They may be used to augment real-world evidence "target trial" studies (\cite{Dickerman2019}, \cite{Challa2020}) but are particularly beneficial in typical high dimensional setting presented by molecular and bioinformatic studies (\cite{Pang2016-xd}, \cite{Ju2019-eb}).   

However, recent applications have also called into question the reliability of such approaches in real-world data (\cite{Yu2019-ah}). To estimating causal effect parameters, the use of doubly-robust (DR) efficient estimators (Augmented Inverse Probability Weighting, AIPW; Targeted Minimum Loss Estimation, TMLE) coupled with ensemble learning of nuisance parameters has attracted much deserved attention (\cite{naimi2021challenges}, \cite{balzer2021comment}, \cite{rotnitzky2019characterization}, \cite{pang2016effect}). The main features of robustness to either treatment or outcome model mis-specification and the ability to overcome slower convergence rates of data-adaptive algorithms (\emph{i.e.} flexible, non-smooth, "machine learning") are key practical features recommending their uptake for observational clinical research settings, where covariate data will be abundant, but model mis-specification (\emph{e.g.} functional forms and/or variable selection) is nearly assured. 

Recent works have highlighted some core considerations for investigators planning to apply these methods. These include the insufficiency of singly-robust methods to achieve bias minimization \cite{naimi2021challenges} and the need for crossfitting, i.e., target parameter and standard error estimation on separate datasets from nuisance function estimation (\cite{newey2018cross}, \cite{zivich2021machine}) to produce valid standard errors, even when treatment or outcome models are slightly  mis-specified. Technically, this occurs when algorithms that are highly data-adaptive do not fulfill Donsker conditions necessary for semi-parametric estimation, resulting in an asymptotically non-negligible empirical process bias. Since such algorithms are difficult to classify, for simplicity, we will refer (somewhat imprecisely) to algorithms that likely violate these conditions, such as random forests or boosted regression trees, as non-smooth, data-adaptive, flexible, or "machine learning," interchangeably. We will refer to algorithms that are likely to suffice, such as penalized linear regression and multivariate polynomial splines, as smooth (differentiable) algorithms. 

More fundamentally, small real-world sample sizes general prevent any guarantees of appropriate coverage properties regardless of the types of estimators implemented (\cite{benkeser2017doubly}, \cite{rotnitzky2019characterization}, \cite{balzer2021comment}). For example,  \cite{benkeser2017doubly} demonstrated an approach for doubly robust inference, \emph{i.e.} nominal asymptotic coverage when using non-smooth estimators. However, performance in small sample ($n<500$) remained sub-optimal. Nonetheless, the implementation of double-crossfit, doubly-robust estimators with sufficiently diverse flexible estimation of nuisance parameters have appeared in these studies to have optimal bias and variance properties amongst possible alternatives even amongst small sample sizes (\cite{balzer2021comment}). As \cite{balzer2021comment} points out, however, these work have generally evaluated estimators on simple data sets that differ in critical ways from typical data and model-fitting procedures applied in clinical and other biomedical contexts. For example, in molecular epidemiologic settings measured covariates will be high-dimensional, while cohort sizes will be moderate due to cost and logistical challenges of follow-up and measurement. Assessing performance in simulated data sets as close to the target context as possible is especially critical (\cite{morris2019}, \cite{boulesteix2017}, \cite{stokes2020}) when complex data present many identification and estimation threats, including conditions not often considered, such as inclusion of inappropriate (\emph{e.g.} near-instruments) or excess covariates irrelevant to the data generating process. Given the major promise of doubly-robust methods in performance under model misspecification, particularly the optimization of nuisance models through data-adaptive ML approaches, properly modeling the target analytic context (\emph{i.e.} data structure with relevant features) is key. Evaluation of these methods against standard approaches in more representative data settings will give a better sense of their potential real-world performance. 

Previous simulation studies have used fairly simple confounding structures (\cite{naimi2021challenges}) and generally large sample sizes (\cite{bahamyirou2019understanding},  \cite{zivich2021machine}) to clearly demonstrate and isolate certain threats to estimators. However, such settings may be overly optimistic in terms of both data structures and model-fitting practices in clinical and molecular epidemiologic settings. In the latter settings, practical concerns such as overadjustment of high-dimension covariates (\emph{i.e.} adjustment of near-instruments), practical positivity violations, small samples, and mis-specification of average treatment effects, \emph{e.g.} by omission of key biological interactions, will be common. Moreover, the typical applied researcher may not be able to spend much time tuning algorithmic hyper-parameter, an important factor even in ensemble learning performance \cite{naimi2021challenges}. One past effort by \cite{bahamyirou2019understanding} focusing on positivity violations in propensity score estimation also only considered large samples and few, simple covariates, and only used an approach to address propensity score fitting (collaborative-TMLE; CTMLE \cite{van2010collaborative}) and not cross-fitting of the overall estimator. Work by Pang, et al (\cite{pang2016effect}) utilized a covariate structure-preserving "plasmode" simulation method (e.g. \cite{franklin2014plasmode}) to evaluate TMLE in the presence of high dimensional covariates. However, the work was conducted in the context of large administrative pharmacoepidemiologic databases (N >16,000) and did not consider non-smooth learning nuisance parameter a key value of efficient estimators. Taken together, these results present an optimistic picture of the robustness of novel estimators that may not apply to all observational health research settings. 

In this paper, we demonstrate performance in closer to real-world conditions for clinical and molecular epidemiologic studies, which could greatly benefit from the efficiency and robustness properties of newer estimators. Notably, we apply efficient estimators (AIPW, TMLE) with ensemble learning of nuisance parameters to estimate average treatment effects under various scenarios of mis-specification. We fit models with and without double-crossfitting and with smooth and/or non-smooth algorithms. Covariate data were drawn from an existing longitudinal cohort study (N = 1178; 331 covariates) to simulate treatment and outcome values under user-specified models ("plasmode" simulation). We then present a tool REFINE2 (A tool to evaluate real-world performance of machine-learning based effect estimators for molecular and clinical studies) which automates this process by taking user input datasets, generating synthetic data with fixed effect sizes, and testing various machine-learning based estimators under user-specified conditions.

The organisation of the paper is as follows: In Section 2, we introduce the estimand of interest and estimators we will be comparing. Section 3 replicates a classic simulation study demonstrating the basic properties of these estimators under model mis-specification, and confirming their good asymptotic performance in basic simulated data. In Section 4, we describe the overall simulation method and the three scenarios we simulated from existing cohort data. In Section 5 we present the results of the analyses and in Section 6 discuss implications and give final recommendations.

\section{Methods}
\label{sec:methods}

\subsection{The Target Parameter}
We use the Rubin counterfactual framework to define the causal estimand: Suppose we observe $n$ i.i.d. data $(O_1, O_2,..., O_n)$. Each observation $O_i$ consists of $(W_i, A_i, Y_i)$, where $Y_i \in \mathbb{R}$  denotes the observed outcome, $A_i \in \{0,1\}$ is a binary random variable representing the treatment received, and $W_i \in \mathcal{W} \subset \mathbb{R}^d$ denotes the covariates of $i$th subject. Furthermore, we denote the counterfactuals as $Y(a), a= 0,1$. Each counterfactual $Y(a)$ representing the outcome received has the patient received the treatment $a$. 

The average  treatment effect (ATE) is  then defined as: 
$$\psi_0= E[Y(1)]  - E[Y(0)]$$

% should W_i be bolded to denote a vector?

To identify the ATE, we also adopt the following causal assumptions: 
\begin{itemize}
    \item Exchangeability: $(Y(1), Y(0)) \bot A | P(A=1|W)$, where W is the vector of all confounding variables
    \item Consistency: If $ A = a$,  then $Y(a) = Y$; that is, the observed value of Y at a is equal to potential outcome of $Y$ had $A$ been set to $a$. 
    \item Stable Unit Treatment Values Assumption (SUTVA): No interference between units. 
    \item Structural positivity holds: $P(X=1|W=w)>0$ and $P(X=0|W=w)>0$ for all units where $w \in \mathcal{W}$
\end{itemize}
Based on these assumptions, we can identify the ATE as
$$\psi_0 = E[E[Y|A=1, W] - E[Y|A=1, W]] $$

\subsection{Propensity Score and IPTW}

A conventional approach to exchangeability is via Inverse Probability of Treatment Weighting (IPTW) to form pseudo populations where treatment status is conditionally independent of measured predictors. Based on \cite{rosenbaum1983central} $(Y(1), Y(0)) \bot A | P(A=1|W)$ under strong exchangeability and positivity. Namely, confounding may be controlled by weighting individual observations by the inverse of their propensity score $g_0(W) = P(A=1|W)$: $1/g_0(W)$ for the treated, $1/(1-g_0(W))$ for the control. This is sometimes referred also as Horvitz-Thompson (HT) estimator \cite{horvitz1952generalization}. It is a consistent estimator of the IP weighted mean as long as the estimated propensity score (PS) model $\hat{g}$ is correctly specified. The IPW estimator of the ATE is then given by:
\begin{align*}
   \hat \psi_n^{IPW}= \frac{1}{n}\sum_{i=1}^n \frac{A_i Y_i}{\hat{g}(W_i)}  - \frac{1}{n} \sum_{i=1}^n \frac{(1 - A_i) Y_i}{1 - \hat{g}(W_i)}
\end{align*}

Highly discriminating propensity scores can lead to extreme weights, so a general approach is to scale or stabilize all weights by introducing the observed treatment prevalence $P(A=1)$ to the numerator. As classification improves, so does the presence of extreme weights, particularly in high dimensional settings. Probabilities that are close to 0 or 1 may result in near violation of the positivity assumption, resulting in bias and instability in ATE and variance estimation. We adopt the standard practice of truncating extreme weights to the upper and lower 5\%-iles, trading off some misspecification to improve performance. Approaches to iterative fitting of high-dimensional propensity scores with smooth or more flexible estimators are possible (\emph{e.g.} \cite{schneeweiss2009high}, \cite{koch2017lasso}, \cite{wyss2018using})

\subsection{g-computation}
A second standard approach is the g-computation estimator, which is motivated directly by the ATE identification form known as the G-formula, $E[E[Y|A=1, W] - E[Y|A=0, W]]$. The ATE is estimated by first fitting the outcome model (OM) for the outcome $Y$ conditional on the treatment variable $A$ and covariates $W$. Then, the predicted outcome value for each observation is computed at the assigned treatment value and the treatment-specific mean computed. This is repeated for the control or comparison treatment value and the ATE is given by the difference in mean outcome for the sample population under the counterfactual treatment values (e.g. treatment and control). More specifically, 

\begin{align*}
   \hat \psi_n^{gcomp}= \frac{1}{n}\sum_{i=1}^n m_1( W_i) - \frac{1}{n} \sum_{i=1}^n m_0( W_i)
\end{align*}

where $m_i(w) = \hat E[Y|X=i, W=w]$. As with IPTW, the OM can be fitted with smoothly or more flexible models, though the latter is not recommended (\cite{naimi2021challenges}). The non-parametric bootstrap is typically used to estimate percentile-based standard errors.

% XM_AUG: Why put efficient here?
\subsection{Doubly-robust, efficient estimators: TMLE and AIPW}
Approaches combining both propensity score (PS) and outcome models (OM) such as Targeted Maximum Likelihood (or Minimum Loss) Estimation (TMLE) and Augmented Inverse Probability Weighting (AIPW) are often termed "doubly-robust" estimators. That is, estimation of counterfactual contrasts such as the ATE are consistent (asymptotically unbiased) as long as at least one of the two models (PS or OM) are correctly specified. In this context, the estimated PS (or clever covariate) or OM-based predictor (or offset) are considered to be "nuisance" parameters in that they are not the primary parameters of interest. Nuisance parameters may be fitted by one or more smooth or flexible, data-adaptive algorithms without substantial alterations to the final estimator. In theory, the latter particularly allows for more complete adjustment for confounding or compensation for model misspecification without added complexity to the final estimator. Importantly, these doubly-robust estimators are necessary to compensate for the relatively slower rates of convergence for flexible estimators (even when models are correctly specified). However, as will be discussed in this paper, these approaches may suffer additional limitations in practical application.  

AIPW, also referred to as a one-step correction estimator, is developed based on g-computation with a mean-zero correction term based on the PS for the asymptotic bias arising from an inconsistent (biased) OM model. Predicted probabilities by PS model and outcome by OM are added together to generate predictions of ATE under each value of covariate. More specifically,
\begin{align*}
    \hat \psi_n^{AIPW} = \frac{1}{n} \sum_{i=1}^n (\frac{A_i Y_i}{\hat g(W_i)} - \frac{A_i - \hat g(W_i)}{\hat g(W_i)} m_1(W_i) ) - \frac{1}{n} \sum_{i=1}^n (\frac{(1-A_i) Y_i}{1-\hat g(W_i)} + \frac{A_i - \hat g(W_i)}{1-\hat g(W_i)} m_0(W_i) )
\end{align*}
Standard errors and confidence intervals can be estimated by from the fitted influence functions (\cite{lunceford2004stratification}) or by bootstrap (\cite{funk2011doubly})

TMLE targets the estimation of ATE by correcting the asymptotic bias of g-computation estimator by adjusting the distribution (\cite{schuler2017targeted}) in a slightly different manner than AIPW. 

First, a initial prediction of the outcome Y (scaled to [0,1] for continuous Y) is fitted by the OM ($m_1$ and $m_0$). Next, a smooth submodel with a single parameter estimated by the PS ("clever covariate") with an offset comprising the initial prediction. Specifically, we denote the clever covariate as:
 $$
H\left(A_{i}, W_{i}\right)=\frac{A_{i}}{\hat g(W_i)}-\frac{1-A_{i}}{1-\hat g(W_i)}
$$ The submodel indexed by $\epsilon$ is obtained by fitting the logistic regression
$$
\operatorname{logit}\left(Y_{i}\right)=\operatorname{logit}\left(m_{A_i}(W_i)\right)+\epsilon H\left(A_{i}, W_{i}\right)
$$

The fitted submodel is then used to compute predicted outcome for each treatment level, and average difference of them produces the estimated ATE.

$$
\operatorname{logit}(m_{a}^{*}(W_i))=\operatorname{logit}(m_{a}^{*}\left(m_a,H\left(a, W_{i}\right), \hat{\epsilon}\right))=\operatorname{logit}\left(m_{a}\left(W_{i} \right)+\hat{\epsilon} H\left(a, W_{i}\right)\right), a=0,1
$$

% $$
% \operatorname{logit}(m_{i}^{*}\left(m_i, \hat{\epsilon}\right))=\operatorname{logit}\left(m_{i}\left(W_{i} \right)+\hat{\epsilon} H\left(A_i, W_{i}\right)\right)
% $$

$$
\hat{\psi}_{T M L E}=\frac{1}{n} \sum_{i}^n (m_{1}^{*}(W_i)  -m_{0}^{*}(W_i)) 
$$

% $$
% \hat{\psi}_{T M L E}=\frac{1}{n} \sum_{i} m_{1}^{*}\left(m_{1}, H\left(1, W_{i} \right), \hat{\epsilon}\right)-\frac{1}{n} \sum_{i} m_{0}^{*}\left(m_{0}, H\left(0, W_{i} \right), \hat{\epsilon}\right)
% $$

Standard errors and confidence intervals can again be estimated from fitted influence functions.

\subsection{Double Cross-Fit (DC)}
\label{sec:DC}
A major challenge to doubly-robust estimators arises when flexible, non-smooth estimators are used to fit high-dimensional nuisance functions when target parameters are of low dimensions, as is often the case with applications of AIPW and TMLE. Notably, the rate of bias reduction in the empirical process term (\cite{benkeser2017doubly}) with increasing sample size is insufficient when non-smooth (specifically non-Donsker class) algorithms are applied, leading to asymptotic bias. To address this, cross-fitting procedures were proposed (\cite{chernozhukov2018double} \cite{newey2018cross}) wherein K-fold splits of the data are used to independently fit nuisance models and the final estimator. This approach is asymptotically unbiased if both OM and PS models are correctly specified, as demonstrated in \cite{zivich2021machine}, and can be further doubly robust to variance estimation with some modification. Regardless, even with correctly specified models, bias and under-coverage are possible in small (\emph{e.g.} N < 2000) sample sizes (\cite{benkeser2017doubly}, \cite{balzer2021comment}). Even with small samples (N = 200), it has been suggested but not extensively evaluated that efficient estimators fit with flexible algorithms may outperform smooth learners (\cite{balzer2021comment}).  
\\
In this paper, we investigate this finite-sample performance more closely in simulated data more congenial to real clinical and molecular epidemiologic settings. We apply the double cross-fit (DC) (\cite{zivich2021machine} \cite{newey2018cross}), namely: We separate the data into three equal size random splits. We estimate the OM on split 1, PS model on split 2, then estimate the ATE using split 3. We then rotate the roles of these splits under the constraint that no split is appointed with the same role twice. This produces three different estimators and three split-specific ATE estimates, which are then summarized by a simple mean. To account for stochastic variations in sample splits, we repeat this split-and-estimate process 5 times and take the median as the final ATE estimate. We also investigate whether number of splits (or number of cross-validation folds for ensemble learners substantially influence results).  

\section{Simulation}

\subsection{Data Generation}
\label{sec:basic}
Our primary aim was to demonstrate the relative performance of old and novel ATE estimators on more realistic data sets than those used in past simulation studies, paying particular attention to the relative improvement of flexible, non-smooth algorithms. To benchmark our estimator setup, we replicate the qualitative results of past simulation studies using a common "hard" data generating process (DGP) whose ATE is difficult to estimate with standard smooth approaches (and therefore flexible estimators are favored). Notably, to evaluate that any difference in performance is not strictly due to finite sample bias, we draw a smaller sample than most previous studies, N = 600 to be more comparable to our real data scenario. In this section, we present the set-up and results of this simulation scenario as a baseline to compare across different methods. This DGP appeared first at \cite{kang2007demystifying} and is frequently used to test estimators for ATE because of the extreme non-linear relationship among covariates in both PS and OM (\cite{ning2020robust}, \cite{benkeser2020nonparametric}, \cite{funk2011doubly}, \cite{naimi2021challenges}). 

The data generating process consists of simulating N = 600 rows/observations of: 
5 confounders $W = (W_1, W_2, W_3, W_4, W_5)$ generated as follows:
\begin{align*}
    W_1 \sim N(0,1), \quad W_2 \sim N(W_1 + 2, 2)\\
    W_3 \sim N(2, |2W_2|), \quad W_4 \sim N(W_2^2 + 2 W_3, |W_1|) \\
    W_5 \sim N(W_3 W_4, |W_2 - W_1|).
\end{align*}.
A binary exposure variable $A$ generated by logistic regression:
\begin{align*}
    P(A=1 |W) = \text{logis} (W_1 + W_2/ 20 + W_3 / 50 + W_4/200 + W_5  / 5000)
\end{align*}

A continuous counterfactual outcome $Y(a), a=0,1$ generated by the following:
\begin{align*}
    Y(a) = 6.6 a + 10 W_1 + 0.5 W_2^2 + 0.66 W_3  + 0.25 W_4 + 0.01 W_3 W_4 + 8 \log (W_5) + N(0,4^2)
\end{align*}
Thus here, the true ATE is fixed to be 6.6. 

Finally, the observed outcome variable $Y$ is generated by: 
\begin{align*}
    Y = A\cdot Y(1) + (1-A) \cdot Y(0)
\end{align*}

\subsection{Estimation}
\label{sec:library}
To estimate the ATE for this dataset, we fit IPTW, g-computation, AIPW (standard and cross-fit), and TMLE (standard and cross-fit) estimators using a linear regression model for the OM $E[Y| A=a, W]=\alpha_0 + \alpha_1 W_1 + ...+\alpha_5 W_5 $ and logistic regression for the PS $\text{logit}(P(A=1|W))= \beta_0 + \beta_1 W_1 +...+\beta_5 W_5$ including as covariates $W_1 ,..., W_5$. Note that absent specifications of covariate interactions and non-linearities, these models would be misspecified. 

We then consider different nuisance parameter estimation methods for $E[Y| A=a, W]$ and $P(A=1|W)$ including:
\begin{itemize}
    \item GLM only
    \item Cross-validated ensemble learner (SuperLearner) \cite{van2007super} with a smooth library: GLM, LASSO, multivariate polynomial splines
    \item Cross-validated ensemble learner (SuperLearner) with a non-smooth library: xgboost and random forest
    
\end{itemize}
For simplicity, for each ATE and nusiance parameter estimator combination, we apply the same estimation approach to both OM and PS (\emph{e.g.} GLM for OM and GLM for PS). For each estimate, we present the absolute bias (Estimate - True), standard error (SE) and 95\% confidence interval coverage. For presentation purposes, absolute bias is scaled by 100 in tables. For example, a bias of 1.0 in tables = 0.01 in absolute bias

%%% NEW_AUG: decide on a standard way to represent bias, it says here the pct bias, which is why I was confused before -- I think the absolute bias (scaled by 100 as necessary) is okay, so I've now edited above
%%%% DONE

\subsection{Result}
\label{sec:simu-result}
%[PRESENT SOME QUANTITATIVE RESULTS FROM THIS SIMULATION]

%%% NEW_AUG I THINK THE TABLE FOR THE KANG & SCHAFER SIMULATION IS STILL MISSING?
%%% DONE

The result here, presented in table \ref{tab:hardsim},  re-affirms that in the case of small sample and limited covariates with non-linear relationships, using non-smooth learners to predict the treatment mechanism (PS) in a singly-robust (IPTW) setting is suboptimal, largely due to over fitting on the propensity score which induces large weights and non-positivity. Moreover, as also shown in \cite{zivich2021machine}, \cite{naimi2021challenges}, efficient estimators fit with non-smooth algorithms suffer without cross-fitting. Despite the relative small sample size cross-fit estimators using non-smooth libraries perform relatively well and comparable to the models fit with smooth algorithms. Interestingly, standard g-computation performance was ideal potentially due to better congeniality to the simulation data generating process and instability of the PS (\cite{robins2007gcomp}, \cite{chatton2020gcomp}). As the focus is on flexible algorithms and efficient estimators rather than of a feature of the simulation, we omit g-computation from subsequent simulation studies.  

In the section below, we evaluate whether such qualitative results hold in more realistic finite sample settings with high-dimensional, correlated covariates and various degrees of model misspecification, including the presence of true biological interaction (effect modifiers). We investigate specifically whether crossfitting and non-smooth approaches, which substantially increase computational time, provide improved performance over singly-fit efficient estimators or more conventional regression-based approaches. 

% shouldn't this Table 1 and Figure 1 go in the Results section?
% alternatively, you should have the basic simulation results (N = 600) as a table or figure here

%%% NEW_AUG: YES STILL DON'T KNOW IF THIS IS THE RESULTS FOR KANG & SCHAFER? THE N IS WRONG AND DOESN'T HAVE MULTIPLE RESULTS (GLM, SMOOTH, NON-SMOOTH) FOR IPW, e.g.
%%% now new result 

	\begin{table}[h!]
		\begin{center}
			\caption[Caption for LOF]{Result from Plasmode simulation on bias, mean squared error (MSE) and 95\% confidence interval (CI) coverage. \footnotemark}
			\bigskip
			\label{tab:hardsim}
			\begin{tabular}{rccccccccccc}
				\toprule
				&       \multicolumn{1}{c}{IPW}  &   \multicolumn{2}{c}	{TMLE } &   \multicolumn{2}{c}	{AIPW }  & \multicolumn{2}{c}{DC-TMLE} &   \multicolumn{2}{c}	{DC-AIPW }\\
					\cmidrule(r){3-4} \cmidrule(l){5-6}  \cmidrule(l){7-8} \cmidrule(l){9-10}
				&  &   \multicolumn{1}{c}{S} & \multicolumn{1}{c}{NS} & \multicolumn{1}{c}{S} & \multicolumn{1}{c}{NS}&   \multicolumn{1}{c}{S} & \multicolumn{1}{c}{NS} & \multicolumn{1}{c}{S} & \multicolumn{1}{c}{NS} \\
				\midrule
				\multicolumn{1}{l}{ }	 & & & \\
Bias ($\times$ 100) & -7.56 & 0.3 & 0.03 & 0.35 & 0.34 & 0.14 & 0.36 & 0.23 & 0.08\\  [2pt] 
 SE ($\times$ 100) & 37.85 & 8 & 7.08 & 8.25 & 7.14 & 9.28 & 9.23 & 9.3 & 9.36\\  [2pt] 
 CI covg.  & 0.99 & 0.92 & 0.87 & 0.94 & 0.92 & 0.96 & 0.95 & 0.95 & 0.98\\  [2pt] 
 MSE  & 0.1 & 0.01 & 0.02 & 0.01 & 0.01 & 0.01 & 0.01 & 0.01 & 0.01\\  [2pt] 
 BVar  & 0.1 & 0.01 & 0.02 & 0.01 & 0.01 & 0.01 & 0.01 & 0.01 & 0.01\\  [2pt] 
				\bottomrule 
			\end{tabular}
		\end{center}
	\end{table}
	
\footnotetext{ Sample size is 1178. Number of bootstrap samples is 100. Bias, SE are the median of those from 100 boostraps. S: Smooth learners: GLM, LASSO, cubic splines; NS: non-smooth learners: xgboost andrandom forest. IPW: inverse probability weighting; TMLE: Targeted Maximum Likelihood; AIPW: Augmented IPW; DC-TMLE: Double Cross-fit TMLE; DC-AIPW: Double Cross-fit: IPW (similarly hereinafter)}

% 	\begin{table}[h!]
% 		\begin{center}
% 			\caption[Caption for LOF]{Result from Plasmode simulation on bias, mean squared error (MSE) and 95\% confidence interval (CI) coverage. Sample size is 1178. Number of bootstrap samples is 100. Bias, SE are the median of those from 100 boostraps \footnotemark}
% 			\bigskip
% 			\label{tab:hardsim}
% 			\begin{tabular}{rccccccccccc}
% 				\toprule
% 				&       \multicolumn{1}{c}{IPW}  &   \multicolumn{2}{c}	{TMLE } &   \multicolumn{2}{c}	{AIPW }  & \multicolumn{2}{c}{DC-TMLE} &   \multicolumn{2}{c}	{DC-AIPW }\\
% 					\cmidrule(r){3-4} \cmidrule(l){5-6}  \cmidrule(l){7-8} \cmidrule(l){9-10}
% 				&  &   \multicolumn{1}{c}{S} & \multicolumn{1}{c}{NS} & \multicolumn{1}{c}{S} & \multicolumn{1}{c}{NS}&   \multicolumn{1}{c}{S} & \multicolumn{1}{c}{NS} & \multicolumn{1}{c}{S} & \multicolumn{1}{c}{NS} \\
% 				\midrule
% 				\multicolumn{1}{l}{ }	 & & & \\
% 				Bias ($\times$ 100) & 11.89 & 1.71 & 13.33 & -0.66 & -4.95 & 2.63 & 2.34 & 2.51 & -7.78\\  [2pt] 
%  SE  & 0.12 & 0.12 & 0.04 & 0.13 & 0.04 & 0.15 & 0.16 & 0.16 & 0.16\\  [2pt] 
%  CI covg.  & 0.82 & 0.8 & 0.28 & 0.92 & 0.45 & 0.95 & 0.97 & 0.96 & 0.94\\  
% 				\bottomrule 
% 			\end{tabular}
% 		\end{center}
% 	\end{table}

%%% NEW_AUG: I THINK I PREFER "S" for smooth and "NS" for non-smooth in the headers & legends
%%% DONE

\section{Plasmode}

\subsection{Source data and plasmode simulation approach}
To generate our data generating process, we borrow covariate data from the Growing Up in Singapore Towards healthy Outcomes (GUSTO) prospective birth cohort (\cite{soh2014gusto}), a population-based, deeply genotyped and phenotyped mother-offspring longitudinal cohort designed to investigate genetic, environmental, and behavioral influences on child and adolescent physical and mental health. The dataset has $n=1178$ observations and covariates of dimension $p=331$ consisting of genetic and other molecular biomarkers (\emph{e.g.} micronutrients), sociodemographic characteristics, behavioral measures, clinical measurements, and medical history data. Variable identities were anonymized as they were not material to the data generating process, though variables plausibly related to effect measure modification (\emph{e.g.} ethnicity and sex were noted for simulation purposes). To simplify the simulation task, a single imputation set by multivariate iterated regressions (\emph{i.e.} chained equations) was taken to form a complete case starting set.  A simple exploratory analysis can be found in appendix.

% The following shows correlation among these covariates:
% \begin{figure}
%     \centering
%     \includegraphics[width=0.9\linewidth]{CovarCorr.png}
%     \caption{Correlation matrix among covariates. Among $\binom{331}{2} = 54615$ pairs of correlation, $30$ of these pairs have absolute value greater than $0.7$, $13$ pairs greater than $0.8$, $4$ pairs greater than $0.9$}
%     \label{fig:covarcor}
% \end{figure}

We specified a scientific task of estimating the ATE of maternal pre-pregnancy obesity status (binary treatment) on child weight in kg (continuous outcome). In scenarios where true biological interaction exists, this ATE was defined as the subgroup-specific effects marginalized to the observed distribution of all effect modifiers. Within the set of covariates we specified five \emph{a priori} demographic, genetic, and maternal comorbidity variables (ethnicity, child sex, gestational diabetes, gestational hypertension, and obesity polygenic risk score) which may plausibly be effect modifiers on the additive scale, along with highly-correlated molecular biomarkers (\emph{e.g.} standarized DNA methylation values) which may be proxies for underlying health conditions as either (near-) instruments or true confounders. The former, if included in outcome models, may amplify biases from unmodelled confounders (\cite{stokes2020}).

These covariates were then used to simulate treatment and outcome values under several PS and OM models, respectively (as specified below), such that the true effect size is know. The value of these approaches is in the ease of retaining a high-dimensional covariate structure from the source data (plasmode simulation; \cite{franklin2014plasmode}). The covariate data are bootstrapped and treatment and outcome values assigned by the approach outlined below. Importantly, overall prevalence of treatment is fixed to that observed in the original data to maintain congeniality with the original dataset in term of covariates that are near-instruments (\cite{stokes2020}). 

%%% JYH: summarize the DGPs and estimators fit above, so the reader knows what to expect in the discussion below.---I think that was answered in section 3.3?

%%% XM: Redo this paragraph later
% JYH: this is redundant now I think
%The main advantage of plasmode is to keep the correlation structure among observed covariates %preserved while we can specify user-chosen true ATE. We believe that this is much closer to real %world setting than made-up simulations when it comes to compare performance of estimators. 

%%% JYH: I would delete / incorporate this paragraph into the text above. It has been stated several times already.---Done

The procedure to generate a single (plasmode) dataset is as follows:
\begin{enumerate}
    \item Specify OM and PS for desired data generating mechanism (specifications for models follow in sections \ref{sec:models})

    \item Fit the PS model using the data. Use the estimated coefficients to re-sample treatment variable, but modifying the intercept of the treatment variable to preserve observed treatment prevalence.
    
    \item Estimate coefficients based on the OM from the whole data. Manually set the main coefficient for the treatment variable to the desired ATE (e.g. 6.6 a plausible increase in child weight due to maternal obesity status). Interaction terms, if involved, remain intact.
    
    \item Generate the outcome using the OM with modified treatment coefficients and add error terms by randomly sampling the residuals of the OM with replacement.
\end{enumerate}

This procedure was repeated on 100 bootstrap samples (for each scenario) sampled with replaced. Several bootstrap sample sizes were evaluated and 100 was chosen as there was only small between-bootstrap sample variation in estimated ATEs (< 0.05 in all cases except for non-smooth singly fit TMLE), within bootstrap variance (BVar) is generally very low ($\leq 0.2$) as shown in \ref{tab:A.est}, \ref{tab:B.est}, \ref{tab:C.est}, and no appreciable change to estimates were observed with more bootstrapped sets. (\cite{koehler2009assessment})

Importantly, because the data generating process is equivalent to the g-computation method without any possibility that fluctuations introduce bias (\emph{i.e.} as in AIPW and TMLE), we exclude standard parametric g-computation as a comparison approach and instead focus on comparisons between different implementations of influence function-based efficient estimators. 

The R code for double cross-fit estimators is adapted and modified from (\cite{zivich2021machine}. We modified the code correspondingly to accommodate our data types. In addition, we found our simulation have stable results among different cross-fit splits (5, 10, 20). Hence, we only take the median of 5 splits as our final ATE estimate, compared to 100 splits in \cite{zivich2021machine}.

%%% JYH: I would say something about the structure of the covariates here -- perhaps a correlation heat map for the sampled coviarates would be good. As well as a qualitative description of the parameters generated through plasmode -- are there some variables with large coefficients and others small, were the variables standardized? A histogram or other visualization of true parameters would be helpful---Done

\subsection{Simulation scenarios}

We consider three data generating scenarios: A - interactions amongst covariates with many covariates, B - mis-specification of interactions amongst covariates with few true covariates, and C - mis-specification of true biological interactions with the treatment. Within each, we consider various degrees of estimation model mis-specification of PS and OM. For each estimation, we evaluate IPW, TMLE, AIPW, crossfit TMLE, and crossfit AIPW. Algorithm selection was identical to Section \ref{sec:library}.  We then apply those PS and OM to estimators introduced in Section \ref{sec:methods}. The true ATE is set as 6.6 for Scenario A and B, and due to the true distribution of effect modifiers, 1.193 for Scenario C (computed by 10000 simulated datasets).

For brevity, we present all models (both data generating models and estimation models) in R formula (R Core \cite{rmodelformulae}) pseudocode instead of the full equations. This allows the reader to more easily identify the difference between modelled scenarios at the cost of fully expressing the nuisance terms. Most importantly, $(X+Y)^2$ denotes the inclusion of both first order terms $X$ and $Y$ as well as their product (interaction term) in the regression model. For greater than two terms, \emph{e.g.} $(X + Y + Z)^2$, this denotes all first order terms and possible pairs of two-way interactions (\emph{e.g.} $X\cdot Y$, $X\cdot Z$, $Y\cdot Z$). 

\subsubsection{Scenario A: Mis-specification of interactions among covariates}
\label{sec:models}
First, we consider the case when there are interactions among some covariates. This is a basic scenario extending the simulation in \ref{sec:basic} where mis-specification of nuisance functions can occur by insufficiently rich models, a limitation that may be addressed by ensemble learning approaches. We extend this to the high-dimensional case in which there may be more or less strongly correlated variables as well as potential near-instruments that could severely bias estimates. 

In this case, the true data generating mechanism (OM and PS, respectively) is given by:
\begin{align*}
    Y \sim   A + ( X_1 + X_2 + X_5 + X_{18} + X_{217} )^2 + X_1 + ... + X_{40} \\
    \text{logit}(P(A=1|X)) \sim ( X_1 + X_2 + X_5 + X_{18} + X_{217} )^2 + X_1 + ... + X_{40} 
\end{align*}

These variables were chosen \emph{a priori} for their potential influence (and empirically verified univariate correlations) on other covariates as well as the treatment and outcome including  demographic variables, clinical comorbidities, and a polygenic risk score for obesity. In scenario C, we further consider these variables and effect modifiers. 

We consider three kinds of model misspecification in estimation: 
\begin{enumerate}
    \item (A.cor) Correct specification for both OM and PS. In this case, we expect all estimators to perform well. 
    \item (A.less.1st)  Misspecification in first order terms: 
    \begin{align*}
        Y \sim A + ( X_1 + X_2 + X_5 + X_{18} + X_{217} )^2 + \sum_{i \in \mathcal{I}_{10}} X_i\\
        \text{logit}(P(A=1|X)) \sim  ( X_1 + X_2 + X_5 + X_{18} + X_{217} )^2 + \sum_{i \in \mathcal{I}_{10}} X_i
    \end{align*}
    where $\mathcal{I}_{10}$  is a random subset of $\{1,2,...,40\}$.  This way we misspecify models only in the first order terms. 
    
    \item (A.no.int) No interaction terms: OM, PS models assume no interaction terms. \begin{align*}
    Y \sim   A +  X_1 + X_2 + X_5 + X_{18} + X_{217}  + X_1 + ... + X_{40} \\
    \text{logit}(P(A=1|X)) \sim  X_1 + X_2 + X_5 + X_{18} + X_{217} + X_1 + ... + X_{40} 
\end{align*}
\end{enumerate}

\subsubsection{Scenario B: Interaction among some  covariates, estimation with more covariates used in data generation}
Second, we investigate estimators in a reverse setting as in scenario A. Recall that for A.less.1st, we misspecify the model by considering fewer first-order covariates than necessary (residual confounding / under-identification). In scenario B, we use fewer true covariates and introduce spurious covariates unrelated to the data generating processes (potential over-identification). Again ensemble learning approaches that incorporate penalization should perform better than standard regression approaches here. Moreover, because the nusiance functions will be provided all relevant covariates (as opposed to Scenario A) there is a chance that doubly robust estimators are unbiased.  

The outcome and treatment are generated by, respectively:
    \begin{align*}
        Y \sim A + ( X_1 + X_2 + X_5 + X_{18} + X_{217} )^2 + \sum_{i \in \mathcal{I}_{10}} X_i\\
        \text{logit}(P(A=1|X)) \sim  ( X_1 + X_2 + X_5 + X_{18} + X_{217} )^2 + \sum_{i \in \mathcal{I}_{10}} X_i
    \end{align*}

The fitted OM and PS for estimation are given by, respectively:
\begin{align*}
    Y \sim   A + ( X_1 + X_2 + X_5 + X_{18} + X_{217} )^2 + X_1 + ... + X_{40} \\
    \text{logit}(P(A=1|X)) \sim ( X_1 + X_2 + X_5 + X_{18} + X_{217} )^2 + X_1 + ... + X_{40} 
\end{align*}
which is the same as used in A.cor

\subsubsection{Scenario C: Interaction between covariate and treatment }
Lastly, we consider a more complicated case, namely to include interaction between covariates and treatment, resulting in a population-specific ATE (weighted by the observed distribution of effect modifiers). This scenario is important in that different estimators may be implicitly target different populations based on how effects are marginalized, with the most obvious example being an IPTW vs. a g-computation approach. In turn, different estimators may be more or less susceptible to misspecification of effect modifiers (\cite{conzuelo2021performance}). As discussed, we chose five a priori covariates as true effect modifiers. 

The data generating mechanism for Scenario C is given by:

    \begin{align*}
        Y \sim (A+ X_1 + X_2 + X_5 + X_{18} + X_{217} )^2 \\
        \text{logit}(P(A=1|X)) \sim  ( X_1 + X_2 + X_5 + X_{18} + X_{217} )^2
    \end{align*}

Note that in this case, while we still fix the coefficient for $A$ to 6.6 in the data generating process,  the marginal ATE will differ due to the presence of treatment-covariate interactions and observed covariate distributions. To boost the effect of interactions we artificially inflate the coefficients for interaction terms by a factor of 5. We calculate the true ATE by g-computation within each of 10000 bootstrap samples of the same size ($n=1178$) and taking their mean. The true ATE in this case is 1.193.
%%% NEW_AUG: I think we need to give some information on the distribution of the true ATE in the 500 bootstraps, people may complain if it's not stable may be some issue with our simulations...
%%% DONE, now changed to 10000

%%% JYH: Didn't we discuss mis-specifying the treatment model as well?  

Estimation model mis-specifications considered here are similar to A:
\begin{enumerate}
    \item (C.cor) Correct specification for both OM and PS. 
    
    \item (C.part) In the partially misspecified case, we omit two of the true interaction terms.
    The estimation model is given by:
    \begin{align*}
        Y \sim A + A:X_1 + A:X_{217} +  (X_1 + X_2 + X_5 + X_{18} + X_{217} )^2 \\
        \text{logit}(P(A=1|X)) \sim  ( X_1 + X_2 + X_5 + X_{18} + X_{217} )^2
    \end{align*}
    
    \item (C.bad) In the most extreme case, we specify no interaction terms between covariate and treatment. 
 
 Note that unlike Scenario A, the PS model is correctly specified in every case to target the same population, so at worst, each model is only singly mis-specified.   
    
\end{enumerate}

\section{Result}

\subsection{Estimation results}

%%% JYH: How did you justify 100 bootstraps? Did you look at the Monte Carlo / simulation error? consider: Am Stat. 2009 May 1; 63(2): 155–162. doi:10.1198/tast.2009.0030 -- explained, but please cite this paper where appropriate--Done

\subsubsection{Result on Scenario A: Interactions among some covariates}

	\begin{table}[h!]
		\begin{center}
			\caption[Caption for LOF]{Scenario A: Result from Plasmode simulation on bias, mean squared error (MSE) and 95\% confidence interval (CI) coverage.  \footnotemark }
			\bigskip
			\label{tab:A.est}
			\begin{tabular}{rccccccccccc}
				\toprule
				&       \multicolumn{1}{c}{IPW}  &   \multicolumn{2}{c}	{TMLE } &   \multicolumn{2}{c}	{AIPW }  & \multicolumn{2}{c}{DC-TMLE} &   \multicolumn{2}{c}	{DC-AIPW }\\
					\cmidrule(r){3-4} \cmidrule(l){5-6}  \cmidrule(l){7-8} \cmidrule(l){9-10}
				&  &   \multicolumn{1}{c}{S} & \multicolumn{1}{c}{NS} & \multicolumn{1}{c}{S} & \multicolumn{1}{c}{NS}&   \multicolumn{1}{c}{S} & \multicolumn{1}{c}{NS} & \multicolumn{1}{c}{S} & \multicolumn{1}{c}{NS} \\
				\midrule
				\multicolumn{1}{l}{A.cor }	 & & & \\
Bias ($\times$ 100) & 36.18 & 4.3 & 9.13 & 0.71 & -7.79 & 9.96 & 24.23 & 1.67 & -15.78\\  [2pt] 
 SE ($\times$ 100) & 17.38 & 8.6 & 3.52 & 19.64 & 3.46 & 16.48 & 15.34 & 14.65 & 12.72\\  [2pt] 
 CI covg.  & 0.5 & 0.45 & 0.29 & 0.87 & 0.32 & 0.79 & 0.57 & 0.89 & 0.72\\  [2pt] 
 MSE  & 0.16 & 0.12 & 1.07 & 0.17 & 0.24 & 0.07 & 0.15 & 0.03 & 0.05\\  [2pt] 
 BVar  & 0.02 & 0.11 & 0.92 & 0.17 & 0.2 & 0.06 & 0.08 & 0.03 & 0.03\\  [10pt] 
				\multicolumn{1}{l}{A.no.int}	 & & & \\
Bias ($\times$ 100) & 34.35 & 5.65 & 14.84 & 2.47 & -4.48 & 9.44 & 26.54 & 1.96 & -16.31\\  [2pt] 
 SE  & 19.4 & 8.67 & 3.55 & 16.16 & 3.5 & 16.26 & 16.07 & 14.85 & 12.94\\  [2pt] 
 CI covg.  & 0.66 & 0.38 & 0.22 & 0.81 & 0.29 & 0.75 & 0.66 & 0.92 & 0.76\\  [2pt] 
 MSE  & 0.14 & 0.12 & 1.35 & 0.12 & 0.21 & 0.08 & 0.12 & 0.03 & 0.05\\  [2pt] 
 BVar  & 0.02 & 0.11 & 1.12 & 0.12 & 0.19 & 0.07 & 0.06 & 0.03 & 0.03\\  [10pt] 
				\multicolumn{1}{l}{A.less.1st }	 & & & \\
Bias ($\times$ 100) & 37.85 & 37.95 & 49.16 & 35.53 & 23.91 & 36.82 & 40.32 & 34.43 & 32.07\\  [2pt] 
 SE  & 22.71 & 11.96 & 3.89 & 12.26 & 3.81 & 15.04 & 14.99 & 15.63 & 15.1\\  [2pt] 
 CI covg.  & 0.62 & 0.18 & 0 & 0.28 & 0.08 & 0.34 & 0.35 & 0.4 & 0.44\\  [2pt] 
 MSE  & 0.17 & 0.2 & 1.18 & 0.13 & 0.19 & 0.14 & 0.17 & 0.13 & 0.12\\  [2pt] 
 BVar  & 0.02 & 0.05 & 0.65 & 0.02 & 0.17 & 0.02 & 0.02 & 0.02 & 0.02\\
				\bottomrule 
			\end{tabular}
		\end{center}
	\end{table}
	
	\footnotetext{Sample size is 1178. Number of bootstrap samples is 100. Bias, SE are the median of those from 100 boostraps. CI covg.: 95\% confidence interval coverage; A.cor: Estimation with correct specification of both PS and OM, A.no.int: Estimation without interaction terms among $X_1$, $X_2$,$X_5$,$X_{18}$,$X_{217}$, A.less.1st: Estimation without 1st order covariates other than $X_1$, $X_2$,$X_5$,$X_{18}$,$X_{217}$; MSE: Mean squared error; BVar: variance between bootstrap estimates. \\
S: Smooth learners: GLM, LASSO, cubic splines; NS: non-smooth learners: xgboost andrandom forest. IPW: inverse probability weighting; TMLE: Targeted Maximum Likelihood; AIPW: Augmented IPW; DC-TMLE: Double Cross-fit TMLE; DC-AIPW: Double Cross-fit: IPW (similarly hereinafter)}
	
%%% NEW_AUG: please add a footnote to this one describing the headers "Both, cvControl-2" etc.
%%% DONE

	\begin{table}[h!]
		\begin{center}
			\caption[Caption for LOF]{Result when combining all algorithms (both smooth and non-smooth) in the SuperLearner \footnotemark}
			\bigskip
			\label{tab:A.sens}
			\begin{tabular}{rcccccccccc}
				\toprule
				&     \multicolumn{1}{c}	{TMLE } &   \multicolumn{1}{c}	{AIPW }  & \multicolumn{1}{c}{DC-TMLE} &   \multicolumn{1}{c}	{DC-AIPW }\\
				\midrule
				\multicolumn{1}{l}{num\_CV=2}	 & & & \\
Bias ($\times$ 100) & 23.43 & -1.54 & 15.98 & -0.84\\  [2pt] 
 SE ($\times$ 100)  & 3.58 & 3.51 & 15.46 & 11.24\\  [2pt] 
 CI covg.  & 0.15 & 0.38 & 0.64 & 0.85\\  [2pt] 
 MSE  & 0.42 & 0.03 & 0.11 & 0.03\\  [2pt] 
 BVar  & 0.3 & 0.02 & 0.08 & 0.03\\  [10pt] 

 \multicolumn{1}{l}{num\_CV=10}	 & & & \\
Bias ($\times$ 100) & 23.43 & -2.14 & 17.5 & -1.83\\  [2pt] 
 SE ($\times$ 100) & 3.57 & 3.52 & 14.45 & 10.97\\  [2pt] 
 CI covg.  & 0.1 & 0.36 & 0.64 & 0.84\\  [2pt] 
 MSE  & 0.3 & 0.03 & 0.1 & 0.03\\  [2pt] 
 BVar  & 0.19 & 0.03 & 0.06 & 0.03\\  [2pt] 
				\bottomrule 
			\end{tabular}
		\end{center}
	\end{table}
	\footnotetext{Both: Both smooth and non-smooth learners. num\_CV: Number of cross validation folds used by superlearner to weight different learners for A.cor.  }

% \multicolumn{1}{l}{Original}	 & & & \\
% Bias ($\times$ 100) &  4.3 & 9.13 & 0.71 & -7.79 & 9.96 & 24.23 & 1.67 & -15.78\\  [2pt] 
%  SE  &  0.09 & 0.04 & 0.2 & 0.03 & 0.16 & 0.15 & 0.15 & 0.13\\  [2pt] 
%  CI covg.   & 0.45 & 0.29 & 0.87 & 0.32 & 0.79 & 0.57 & 0.89 & 0.72\\  [2pt] 
%  BVar   & 0.11 & 0.92 & 0.17 & 0.2 & 0.06 & 0.08 & 0.03 & 0.03\\   [10pt] 
				
% 				\multicolumn{1}{l}{All 331 covars}	 & & & \\
% 				Bias ($\times$ 100) & 17.94 & 35.01 & -3.88 & -16.16 & 11.8 & 37.09 & -5.49 & -15.71\\  [2pt] 
%  SE  & 0.03 & 0.03 & 0.03 & 0.03 & 0.16 & 0.16 & 0.14 & 0.13\\  [2pt] 
%  CI covg.  & 0.11 & 0.02 & 0.38 & 0.19 & 0.85 & 0.41 & 0.95 & 0.81\\  [2pt] 
%  BVar  & 0.17 & 1.22 & 0.05 & 0.32 & 0.03 & 0.04 & 0.02 & 0.02\\  [10pt]

In Scenario A where we specify some interactions between predictors of treatment and outcome, even with correctly specified models, every estimator exhibited some sub-optimal performance, particularly in confidence interval coverage (Table \ref{tab:A.est}). AIPW fit with smooth learners (with and without crossfitting) has minimal bias in ATE estimation but coverage did not exceed 89\%. ATE bias was reasonable for TMLE fit with smooth learners regardless of crossfitting (~1-2\%), however coverage was generally poor. Use of non-smooth learners resulted in uniformly poor coverage, though in non-crossfit AIPW/TMLE this was due to overly small SE estimates, while both bias in ATE (~3-4\%) and variance estimation contributed to the doubly-crossfit estimators. Results from models omitting covariate interactions (A.no.int) had comparable performance to the correct model. When covariates were omitted, bias increased across all estimators (~3-8\%) and coverage was uniformly poor. As expected, however, singly-robust IPW had worst performance with all estimation strategies, even when the PS model was correctly specified. 
\\
We conducted sensitivity analyses adding all candidate learners (both smooth and non-smooth) to the ensemble estimation of nuisance functions in Table \ref{tab:A.sens} and found in nearly all cases (even when increasing the number of cross-validation folds for the SuperLeaner) the results of this richer specification to fall in between the smooth and non-smooth strategies. Suprisingly, a non-crossfit TMLE performed worse when all learners were used versus only non-smooth learners (bias: 3.5\% from 1.4\%). The original model using only smooth learners performed better in bias and coverage in every instance but one. For DC-AIPW, using all learners improved bias somewhat (0.1\% from 0.2\%), but with a slight decrease in coverage (85\% from 89\%). When we modified our true data generating process to include all 331 covariates, results differed slightly: Estimators using non-smooth learners still performed worse than smooth learners in every case. However, there was a clear advantage to double-crossfitting with both DC-TMLE and DC-AIPW having reasonable bias and coverage (1.8\%, 85\% and 0.8\%, 95\%, respectively). Increasing the number of double-crossfit procedures (Table \ref{tab:A.cfnum}) from 5 to 10 or 20 did not substantially change the relative performance of any of the estimators, though of course increasing run time proportionally.

\begin{figure}[ht]
    \centering
    \includegraphics[width=0.99\linewidth]{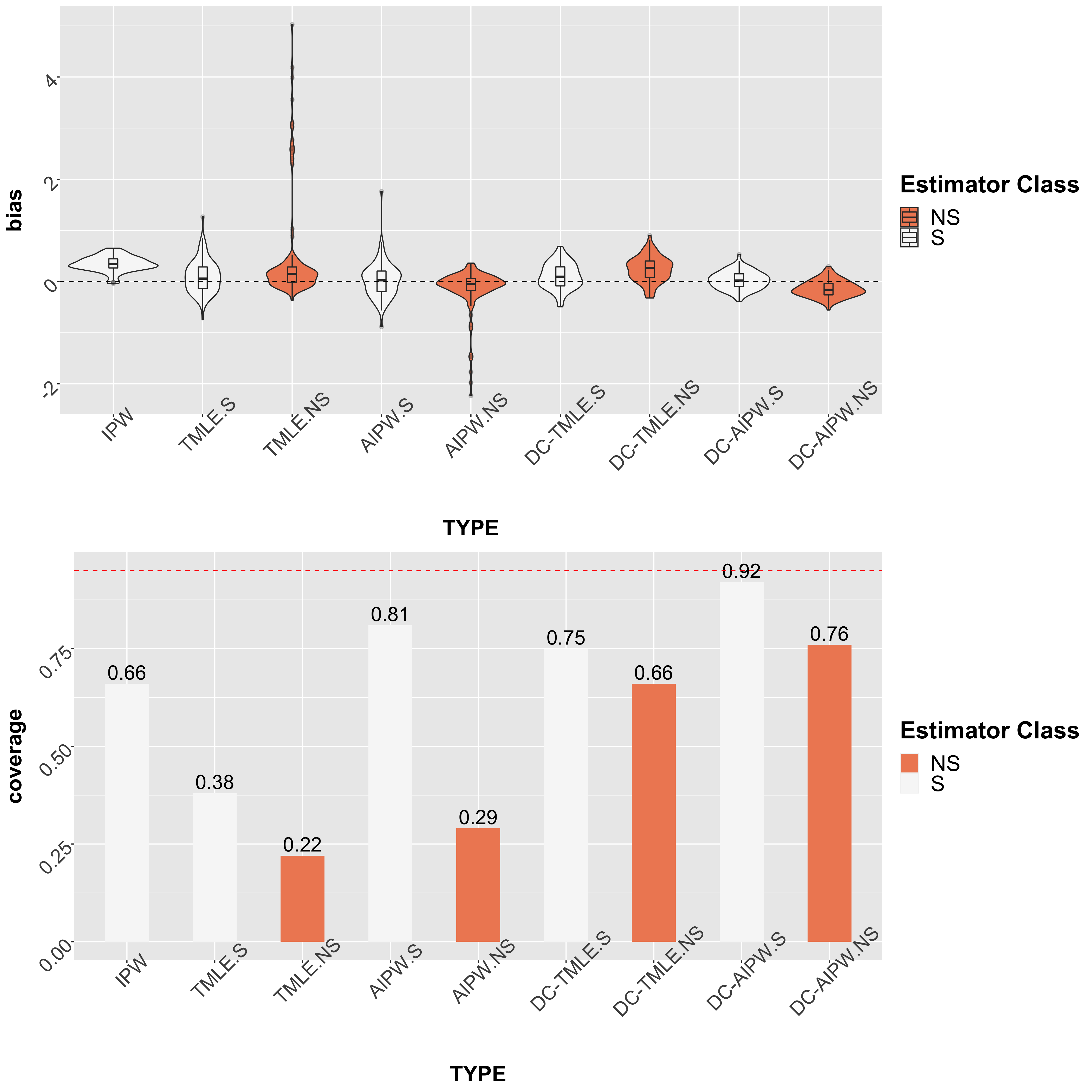}
    \caption{Bias and coverage for all estimators in scenario A.no.int }
    \label{fig:A}
\end{figure}

\subsubsection{Result on Scenario B}

In Scenario B, our true data generating process has fewer variables than our estimation models. Compared to estimators with the correctly specified models, estimators with excess covariates performed slighty worse in terms of coverage, and in the case of non-smooth learners, ATE bias as well (Table \ref{tab:B.est}). Nonetheless, all estimators with smoothly estimated nuisance functions had similarly low bias and reasonable standard error estimates, with crossfit versions having >95\% coverage when excess covariates were included. IPW and non-smoothly estimated TMLE exhibited the worst bias. IPW reached 98\% coverage at the cost of overly large SEs and thus wide confidence intervals. In doubly-crossfit models, non-smooth learners were able to produce similar bias and SE estimates as their smooth counterparts.

%JYH: This table for Scenario B should compare B to B correct (which has 10 covariates) and not A correct (which has 40 covariates). You already showed above there is large differences in 10 vs 40 variables.

%%% JYH: also, can you force the table / figure placement so they appear in the correct part of the text?

%%% NEW_AUG: this info needs to be in a footnote instead: Number of bootstrap samples is 100. Bias, SE are the median of those from 100 bootstraps.
%%% DONE
	\begin{table}[h!]
		\begin{center}
			\caption[Caption for LOF]{Scenario B: Result from Plasmode simulation on bias, mean squared error (MSE) and 95\% confidence interval (CI) coverage. (N = 1178). \footnotemark}
			\bigskip
			\label{tab:B.est}
			\begin{tabular}{rccccccccccc}
				\toprule
				&       \multicolumn{1}{c}{IPW}  &   \multicolumn{2}{c}	{TMLE } &   \multicolumn{2}{c}	{AIPW }  & \multicolumn{2}{c}{DC-TMLE} &   \multicolumn{2}{c}	{DC-AIPW }\\
					\cmidrule(r){3-4} \cmidrule(l){5-6}  \cmidrule(l){7-8} \cmidrule(l){9-10}
				&  &   \multicolumn{1}{c}{S} & \multicolumn{1}{c}{NS} & \multicolumn{1}{c}{S} & \multicolumn{1}{c}{NS}&   \multicolumn{1}{c}{S} & \multicolumn{1}{c}{NS} & \multicolumn{1}{c}{S} & \multicolumn{1}{c}{NS} \\
				\midrule
					\multicolumn{1}{l}{B.cor }	 & & & \\
Bias ($\times$ 100) & 11.72 & -0.51 & 4.53 & 1.62 & -2.63 & 2.54 & -0.97 & 4.13 & -1.38\\  [2pt] 
 SE ($\times$ 100) & 23.18 & 12.99 & 3.97 & 13.77 & 3.94 & 15.71 & 15.44 & 16.78 & 15.86\\  [2pt] 
 CI covg.  & 0.98 & 0.9 & 0.41 & 0.95 & 0.48 & 0.96 & 0.93 & 0.95 & 0.97\\  [2pt] 
 MSE  & 0.03 & 0.03 & 0.2 & 0.03 & 0.02 & 0.02 & 0.02 & 0.03 & 0.02\\  [2pt] 
 BVar  & 0.02 & 0.03 & 0.19 & 0.03 & 0.02 & 0.02 & 0.02 & 0.03 & 0.02\\  [10pt] 
				\multicolumn{1}{l}{B}	 & & & \\
Bias ($\times$ 100) & 11.89 & 0.31 & 9.35 & -0.66 & -4.95 & 2.63 & 2.34 & 2.51 & -7.78\\  [2pt] 
 SE ($\times$ 100)  & 24.78 & 12.08 & 3.9 & 13.32 & 3.86 & 15.31 & 15.99 & 16.18 & 15.8\\  [2pt] 
 CI covg.  & 0.99 & 0.8 & 0.38 & 0.92 & 0.45 & 0.95 & 0.97 & 0.96 & 0.94\\  [2pt] 
 MSE  & 0.03 & 0.03 & 0.53 & 0.03 & 0.02 & 0.02 & 0.02 & 0.02 & 0.03\\  [2pt] 
 BVar  & 0.02 & 0.03 & 0.45 & 0.03 & 0.01 & 0.02 & 0.02 & 0.02 & 0.02\\ 
				\bottomrule 
			\end{tabular}
		\end{center}
	\end{table}
\footnotetext{B.cor: Correct specification of both OM and PS. B: Estimation by adding more covariates to both OM and PS. }

\begin{figure}[ht]
    \centering
    \includegraphics[width=0.99\linewidth]{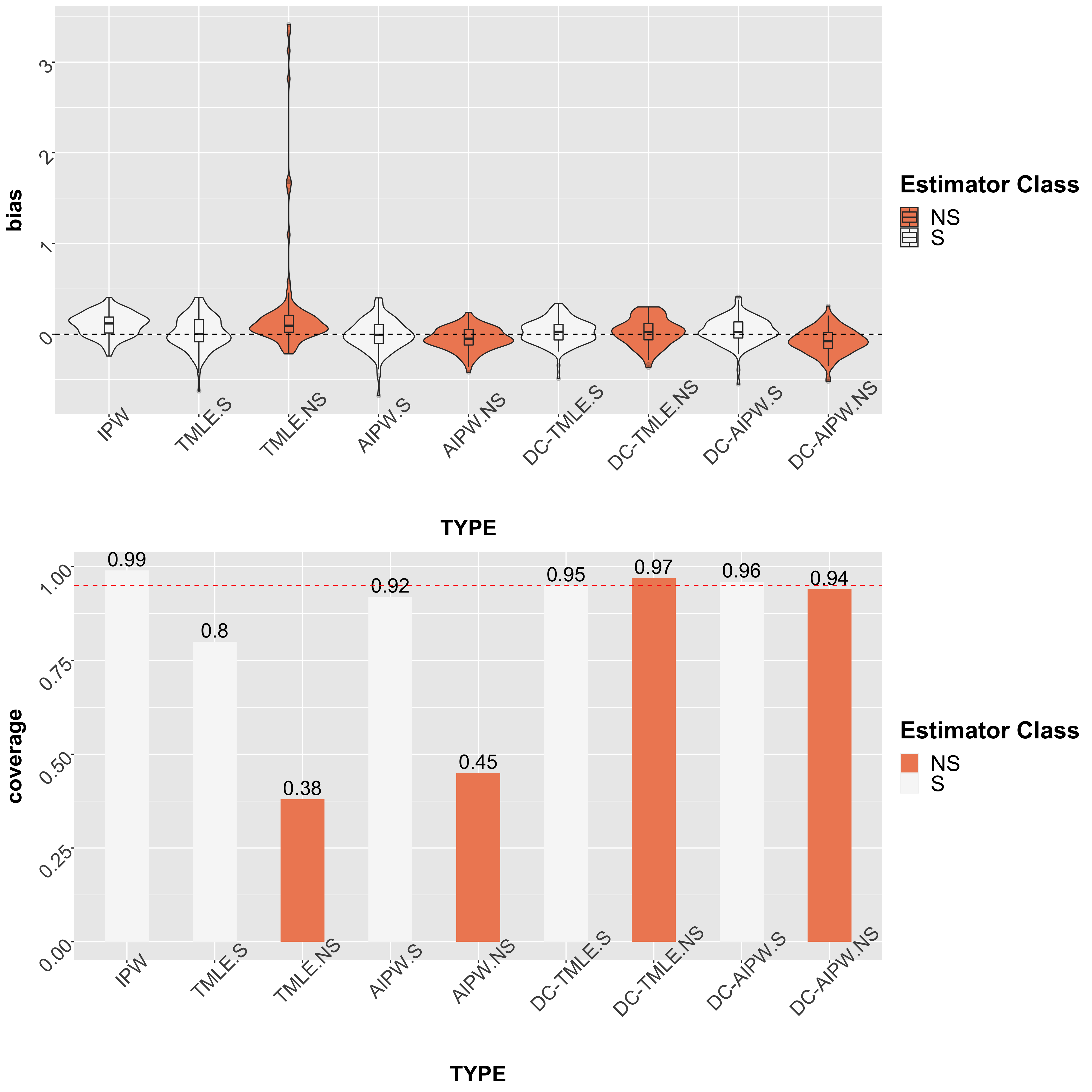}
    \caption{Bias for all 3 cases in scenario B. }
    \label{fig:B}
\end{figure}

\begin{figure}[ht]
    \centering
    \includegraphics[width=0.99\linewidth]{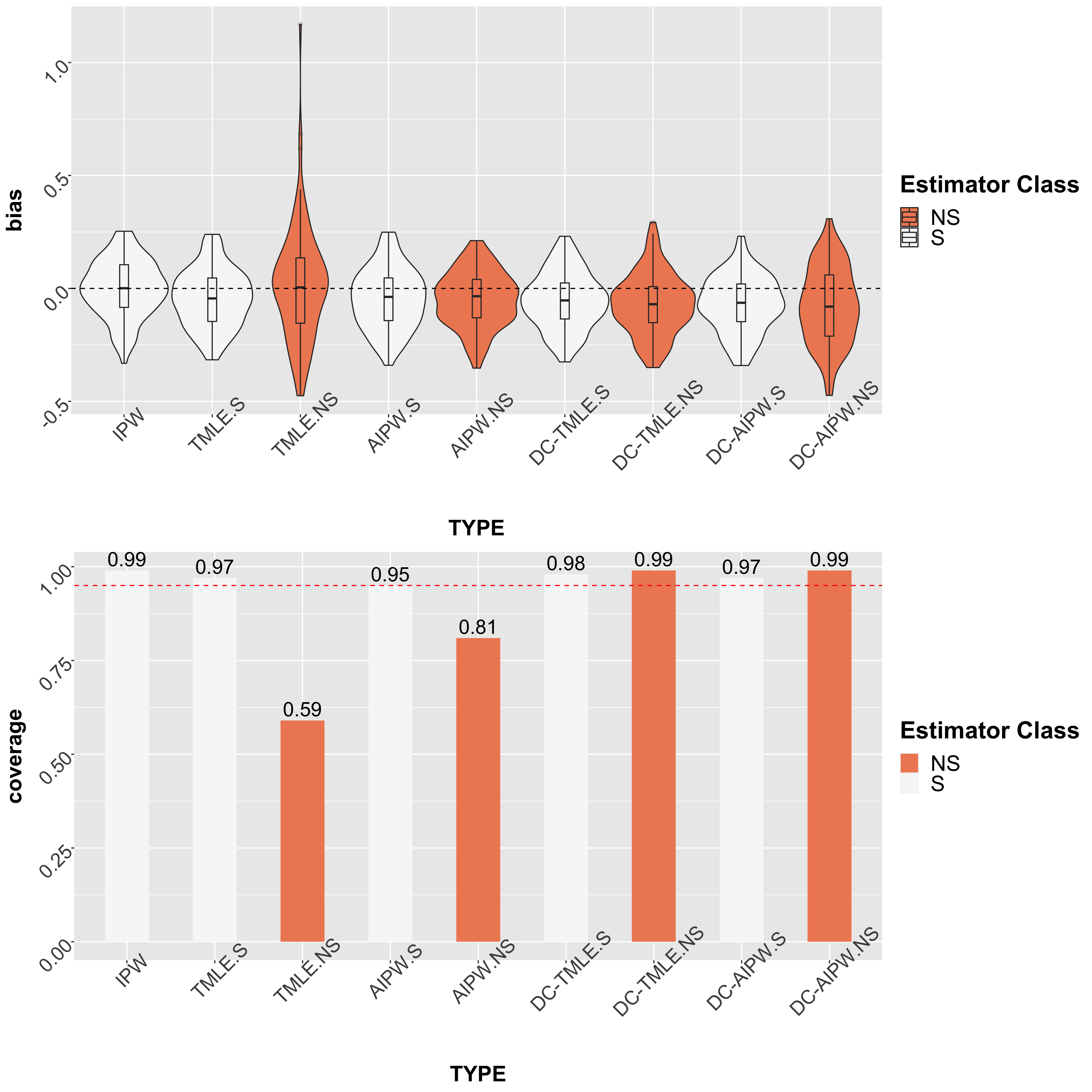}
    \caption{Bias for all 3 cases in scenario C.bad }
    \label{fig:C}
\end{figure}

\subsubsection{Result on Scenario C}
	\begin{table}[h!]
		\begin{center}
			\caption[Caption for LOF]{Scenario C: Result from Plasmode simulation on bias, mean squared error (MSE) and 95\% confidence interval (CI) coverage.  \footnotemark}
			\bigskip
			\label{tab:C.est}
			\begin{tabular}{rcccccccccccc}
				\toprule
				&    \multicolumn{2}{c}	{TMLE } &   \multicolumn{2}{c}	{AIPW }  & \multicolumn{2}{c}{DC-TMLE} &   \multicolumn{2}{c}	{DC-AIPW }\\
					\cmidrule(r){2-3} \cmidrule(l){4-5}  \cmidrule(l){6-7} \cmidrule(l){8-9}
				&    \multicolumn{1}{c}{S} & \multicolumn{1}{c}{NS} & \multicolumn{1}{c}{S} & \multicolumn{1}{c}{NS}&   \multicolumn{1}{c}{S} & \multicolumn{1}{c}{NS} & \multicolumn{1}{c}{S} & \multicolumn{1}{c}{NS} \\
				\midrule
				\multicolumn{1}{l}{C.bad }	 & & & \\
Bias ($\times$ 100) & -4.43 & 0.45 & -3.75 & -3.46 & -5.31 & -6.99 & -6.36 & -8.06\\  [2pt] 
 SE ($\times$ 100) & 13.44 & 8.34 & 13.52 & 8.27 & 15.41 & 20.76 & 16.29 & 21.83\\  [2pt] 
 CI covg.  & 0.97 & 0.59 & 0.95 & 0.81 & 0.98 & 0.99 & 0.97 & 0.99\\  [2pt] 
 MSE  & 0.02 & 0.06 & 0.02 & 0.02 & 0.02 & 0.02 & 0.02 & 0.03\\  [2pt] 
 BVar  & 0.02 & 0.06 & 0.02 & 0.02 & 0.02 & 0.02 & 0.02 & 0.03\\  [10pt] 
				\multicolumn{1}{l}{C.part }	 & & & \\
Bias ($\times$ 100) & -2.76 & 4.12 & -2.53 & 1.2 & -4.28 & -2.5 & -3.74 & -5.88\\  [2pt] 
 SE ($\times$ 100) & 12.72 & 5.77 & 12.86 & 5.77 & 16.51 & 22.51 & 17.43 & 23.97\\  [2pt] 
 CI covg.  & 0.93 & 0.45 & 0.96 & 0.63 & 0.98 & 1 & 0.99 & 0.99\\  [2pt] 
 MSE  & 0.02 & 0.13 & 0.02 & 0.02 & 0.02 & 0.02 & 0.02 & 0.04\\  [2pt] 
 BVar  & 0.02 & 0.12 & 0.02 & 0.02 & 0.01 & 0.02 & 0.02 & 0.03\\  [10pt]
				\multicolumn{1}{l}{C.cor}	 & & & \\
Bias ($\times$ 100) & 0.25 & 0.87 & -0.54 & -1.45 & -2.65 & -3.39 & -2.71 & -0.41\\  [2pt] 
 SE ($\times$ 100) & 11.36 & 4.77 & 11.34 & 4.81 & 12.8 & 17.69 & 13.07 & 22.39\\  [2pt] 
 CI covg.  & 0.94 & 0.38 & 0.93 & 0.49 & 0.97 & 1 & 0.97 & 0.98\\  [2pt] 
 MSE  & 0.02 & 0.08 & 0.02 & 0.02 & 0.01 & 0.02 & 0.02 & 0.03\\  [2pt] 
 BVar  & 0.02 & 0.08 & 0.02 & 0.02 & 0.01 & 0.01 & 0.02 & 0.03\\  
				\bottomrule 
			\end{tabular}
		\end{center}
	\end{table}
\footnotetext{C.bad: Estimate without all interaction terms between covariates and the exposure variable. C.part: Estimate without 3 out of 5 interaction terms between covariates and the exposure variable. C.cor: Estimate with correct specification of PS and OM. }

In scenario C, we fit estimators omitting some or all true interactions (effect measure modifiers) between treatment and five key covariates. Unlike other scenarios, the PS model is correctly specified for every estimator. We find that generally, most estimators performed similarly well (~1-2\% bias) with only non-smooth learner estimated, non-crossfit TMLE and AIPW have higher bias (5-6\%) and poor coverage (between 38\% and 75\%). Smooth learner estimated, non-crossfit TMLE and AIPW had nominal coverage, whereas double-crossfit smooth and non-smooth models were overly conservative (coverage 96\% to 100\%).

\subsection{Timing of algorithms}
In this section, we present the median real time taken for 100 bootstrap samples of each estimator computed on a  Lenovo SD650 NeXtScale server using an Intel 8268 “Cascade Lake”processor and 192GB RAM (table \ref{tab:timing}). As expected, simple IPW estimators take a negligible amount of time to fit. Estimators fit with non-smooth learners were took 2-3 times as long as their smooth counterparts. Crossfit efficient estimators take proportionally more time than their singly fit counterparts, between 3-6 times longer, due to multiple calls to SuperLearner and additional bootstrapping. While it would be possible to parallelize the crossfitting, the computation time for dependent calls to SuperLeaner within each step would not be able to be reduced.

% TODO: run time for one ATE computation for each scenario.

% 	\begin{table}[h!]
% 		\begin{center}
% 			\caption{Timing of each algorithm on 8 cores vs 30 cores  Bootstrap sample size is  600. Number of boostrap samples is 100}
% 			\bigskip
% 			\label{tab:est}
% 			\begin{tabular}{rccccccccccc}
% 				\toprule
% 				&       \multicolumn{1}{c}	{ IPW} &  \multicolumn{1}{c}	{LASSO IPW } &   \multicolumn{2}{c}	{AIPW } & \multicolumn{2}{c}{DC-AIPW} 	&\multicolumn{2}{c}{TMLE } & \multicolumn{2}{c}{DC-TMLE}\\
% 					\cmidrule(r){4-5} \cmidrule(l){6-7} \cmidrule(r){8-9}
% 					\cmidrule(l){10-11}
% 				&  &  &   \multicolumn{1}{c}{Par} & \multicolumn{1}{c}{Non-Par} & \multicolumn{1}{c}{Par} & \multicolumn{1}{c}{Non-Par} &  \multicolumn{1}{c}{Par} & \multicolumn{1}{c}{Non-Par} & \multicolumn{1}{c}{Par} & \multicolumn{1}{c}{Non-Par} \\
% 				\midrule
% 				\multicolumn{1}{l}{A.cor }	 & & & \\
% 				8 cores & & & & & & &  319.64\\  [2pt]
% 				30 cores & & & & & & & 128.72
% 				\\  
% 				\bottomrule 
% 			\end{tabular}
% 		\end{center}
% 	\end{table}

%%% NEW_AUG: Complete and move to footnote: Sample size is 600, and the result is the median of... 
	\begin{table}[h!]
		\begin{center}
			\caption{Timing of each algorithm in seconds.}
			\bigskip
			\label{tab:timing}
			\begin{tabular}{rccccccccccc}
				\toprule
				&       \multicolumn{1}{c}{IPW}  &   \multicolumn{2}{c}	{TMLE } &   \multicolumn{2}{c}	{AIPW }  & \multicolumn{2}{c}{DC-TMLE} &   \multicolumn{2}{c}	{DC-AIPW }\\
					\cmidrule(r){3-4} \cmidrule(l){5-6}  \cmidrule(l){7-8} \cmidrule(l){9-10}
				&  &   \multicolumn{1}{c}{Par} & \multicolumn{1}{c}{Non-Par} & \multicolumn{1}{c}{Par} & \multicolumn{1}{c}{Non-Par}&   \multicolumn{1}{c}{Par} & \multicolumn{1}{c}{Non-Par} & \multicolumn{1}{c}{Par} & \multicolumn{1}{c}{Non-Par} \\
				\midrule
				% \multicolumn{1}{l}{A.cor (w.)}	 & & & \\
				%  & 0.073&   2.993&   27.854&  62.186&  77.183&  128.501& 18.389&  51.295&  81.624&  126.353\\  [2pt]
			 %& 0.069&  2.91&   25.83&  61.877& 65.531& 97.441& 18.121& 51.636& 68.149& 97.299& \\  [10pt]
				\multicolumn{1}{l}{A.cor }	
				  & 0.12  & 20.82 & 55.86 & 27.17 & 63.99 & 117.47 & 204.21 & 120.76 & 210.24\\  [2pt] 
				  \multicolumn{1}{l}{A.less.1st}
				   & 0.12  & 10.13 & 26.2 & 13.56 & 35.95 & 55.01 & 149.86 & 54.68 & 146.55\\  [2pt] 
				   \multicolumn{1}{l}{A.no.int}	
				   & 0.11  & 21.49 & 55.58 & 21.4 & 55.69 & 121.29 & 203.76 & 122.34 & 209.85\\  [2pt] 
				 \multicolumn{1}{l}{C.bad }	 
				  & 0.09  & 3.24 & 11.01 & 5.05 & 27.97 & 18.39 & 72.53 & 17.77 & 71.81\\  [2pt] 
				  \multicolumn{1}{l}{C.part}	
				   & 0.09 & 8.67 & 26.74 & 12.01 & 84.82 & 49.01 & 125.35 & 48.8 & 124.95\\  [2pt] 
				\multicolumn{1}{l}{C.cor}	
				  & 0.12  & 3.45 & 11.17 & 12.34 & 89.81 & 18.34 & 72.33 & 18.05 & 69.91\\  [2pt] 
				\bottomrule 
			\end{tabular}
		\end{center}
	\end{table}

\section{Discussion}
\subsection{Review of the Findings}

Doubly-robust efficient estimators such as AIPW and TMLE represent a state-of-the-art in model-based estimation of causal effects with non-randomized data. Past evaluations of such estimators have been conducted under simpler, more ideal settings \emph{e.g.} low dimensionality, large sample size, appropriate covariate sets (\emph{i.e.} no near-instruments), and correct model specification. As highlighted by \cite{balzer2021comment} evaluation of performance will need to be conducted in each setting they are to be applied. Here, we aimed to investigate some more challenging conditions commonly faced by clinical investigators particularly in molecular epidemiology, namely large covariate sets including the potential for weakly-correlated elements or near-instruments, small sample sizes, and substantial nuisance model misspecifiation. We find in our set-up case mirroring the data generating process posed by \cite{kang2007demystifying}, double-crossfit estimators fit with non-smooth models performed optimally, but notably there was only a drop in performance on bias and coverage by several percent with smooth models and, as expected, only poor estimation of standard errors without cross-fitting for non-smooth models. In contrast, performance in sets simulated from real data showed much poorer performance even in cases where models were correctly specified. However, in nearly every case double-crossfit AIPW/TMLE fit with smooth models were optimal amongst tested algorithms, even when smooth and non-smooth learners were combined in the SuperLearner. 

%%% NEW_AUG: going to remove every mention of the g-computation results, unless you specifically want to discuss the Robins commentary for why it is better
In scenario A, we presented two forms of model misspecification: first, we completely omitted relevant confounders in both treatment and outcome models (A.less.1st); second, we included all covariates but failed to specify interactions among them (A.no.int). We found all models were suboptimal in parameter and variance estimation, even in the correctly specified case. However, both non-crossfit and doubly crossfit AIPW was the closest to unbiased and nominal coverage, reproducing the result of \cite{naimi2021challenges} in the crossfit case. In general, we found that all estimators performed poorly when covariates were omitted (A.less.1st) and less poorly when only covariate-interactions were misspecified. This is likely due to the relatively small contribution of interactions to each nuisance model, but also reflects the ability of SuperLearner to some recover some model misspecification thanks to data-adaptive algorithms. More importantly, however, the smoothly fit estimators performed better than their non-smoothly fit counterparts.  

% since we have shown it again, I think we can make a slightly stronger case than the below:
%Besides, we notice that AIPW has much better performance in A.cor. This %performance difference between AIPW and TMLE in small sample was %demonstrated in  \cite{naimi2017challenges}, and like aformentioned %researchers, we would not make any conclusion on properite between TMLE %and AIPW. Rather, we only recommend the use of smooth models in both %TMLE and AIPW under small sample scenario.
In Scenario B, we test a scenario that has not previously been evaluated with respect to these estimators, but is very common in practice -- namely when models are fit with an excess of variables which are not part of the true data generating process, but which may be correlated to treatment (instrument), outcome, or both by chance. Mathematically, this over-identification should be irrelevant if all true predictors are included, but the presence of slight misspecification of either model may lead to biased estimates \cite{chatton2020gcomp}, notably if chance near-instruments are included in propensity score models \cite{stokes2020} or non-smooth classifiers overfit either model (\cite{balzer2021comment},\cite{bahamyirou2019understanding}). On the other hand, penalized regression such as the LASSO are particularly suited to eliminate weakly correlated predictors. In fact, we find a conventionally fit IPW to be biased and over-cover (1.8\%; 99\% coverage) supporting the substantial literature on bias due to overfit propensity scores. Surprisingly however, doubly-robust estimators fit with non-smooth estimators were also biased and undercover in this scenario (TMLE 1.4\% bias; 38\% coverage) despite having only true confounders or random variables in models. Smooth TMLE and AIPW were nearly unbiased (<0.1\%) with double-crossfit counterparts having nominal coverage at the cost of some slight additional bias (~0.4\%). This suggests all estimators may be quite sensitive to even mild model mis-specifications, in particular, non-smooth-based estimators.

In scenario C, we test the sensitivity of estimators to mis-specification of covariate-treatment interactions, as would be relevant to estimate a population-specific average causal effect (\cite{conzuelo2021performance}). In general, we find that most estimators are in fact relatively unbiased related to other scenarios. However, double-crossfit standard errors were excessively conservative (96\% to 100\%). Interestingly singly-fit TMLE and AIPW appeared to have lowest bias and nominal coverage across all scenarios while, as expected, the same estimators fit with non-smooth learners had generally poorer performance. Notably, there did not appear to be substantial difference in performance across the scenarios. This may be attributed to the fact that propensity score models were corrected specified in every scenario, highlighting the importance to evaluate realistic scenarios where both models are misspecified. 

%%%[In the future, we will consider the influence of mispecification of the propensity score model, though we  do not anticipate changes to our qualitative findings.]  
Overall, we found bias and coverage to be worse across our scenarios and estimators than most past studies, which we attribute mainly to the simulation conditions. Scenarios B and C present estimation problems that are slightly different than past studies, but the relative performance of estimators are reasonably consistent throughout all three scenarios. It is worth highlighting that the fact that bias arises from the chance inclusion of near-instruments further reinforces the need for not only the careful selection of covariates (\cite{chatton2020gcomp}) regardless of the inclusion of penalization algorithms, but also more robust simulation of data congenial to the estimation task faced by the analyst as highlighted by many (\cite{boulesteix2017}, \cite{morris2019}, \cite{stokes2020}, \cite{balzer2021comment}). This work is mostly closely related to \cite{bahamyirou2019understanding} which employed a Bootstrap Diagnostic Test similar in spirit to plasmode simulation (taking the observed treatment coefficient, rather than setting its value) and \cite{pang2016effect} which employed plasmode simulation. However, in the previous cases, larger sample sizes were employed, fewer covariates (in the case of Bahamyirou, et al), and less severe misspecification. Newer doubly-crossfit estimators were also not employed. In every case, including more recent comparisons (\cite{naimi2021challenges},\cite{zivich2021machine}), extent of bias and poor variance estimation appeared to be correctable by improved model specification, hyperparameter tuning, and crossfitting. \cite{benkeser2017doubly} and \cite{balzer2021comment} show that these challenges occur with simple data generating processes in smaller samples (N < 500), but we show them to persist even in larger datasets (N > 1000). More importantly, we show that challenges in more realistic data sets are difficult to overcome even with correctly specified models. However, in typical clinical or molecular epidemiologic applications, double-crossfit models fit with smooth learners may lead to best parameter and variance estimation among possible options.

% may not be necessarily because we compare throughout %\subsection{Past Literature}
%Compared to previous works by \cite{naimi2017challenges} and  \cite{zivich2020machine}, we leveraged on Plasmode to construct bootstrap datasets based on real world data. Besides showing similar results (smooth DR estimators perform better than non-smooth ones in small samples, Double Cross-fit makes inference more reliable) in these Plasmode datasets, we presented cases where Double Cross-Fit can even fail to produce meaningful inference. Moreover, we explicitly showed time taken for each algorithm, addressing another advantage of using simple smooth models over non-smooth models, and over DC estimators. 

\subsection{Limitations}

First, although with Plasmode we are able to retain correlation among other covariates, the treatment distribution (PS) and the outcome model are still parametrically specified. Thus, there is a major concern that models closer to the data generating process are unfairly favored, potentially explaining the slightly better performance of smoothly-fitted AIPW, as highlighted by \cite{naimi2021challenges},  which directly employs the linear g-computation estimate rather than the scaled logistic estimator. To address this, we also omitted from comparisons the parametric g-computation results which are identical to the plasmode simulation process. The dilemma between simulating with a user specified true parameter and retaining real world data structure remains a natural challenge on simulation method design and is a subject for future development.  
Second, there are other important features of data generating processes and estimation strategies that we were not able to test for this study. Notably, while we allowed irrelevant covariates to be included in Scenario B, we did not specifically introduce the effects of strong or near-instruments in models, a challenge which has been demonstrated in other contexts (\cite{stokes2020}). We also did not introduce large libraries or consider extensive hyperparameter tuning as a) this would have substantially contributed to computation time, but more importantly b) we wanted to demonstrate the performance under realistic conditions where the typical clinical science analyst will not spend excessive time on model tuning. 

%Second, we note that we have just came up with three different scenarios showing possible pitfalls on careless usage of estimators. It is interesting to come up with similar scenario on different dataset to show the robustness of our conclusions, and design different batch of scenarios to extend the discussion.

% 1. more advanced methods may not give a better/more credible results.

%We summarized key takeaway messages in this paper as follows: First, made-up simulations are too simple to expose challenges in estimations. Plasmode, although has limitation on , is closer to the real world an. It should be promoted. Second, although non-smooth estimators has flexibility, it can lead to problems such as  overfitting to SE, thus result in unreliable inference. 

\subsection{REFINE2 Tool}
\emph{What is it?}
In this study, we found relative performance of efficient estimators to differ substantially from previous simulation studies and we attribute this to the structure of our specific molecular epidemiologic data at hand. As noted previously \cite{balzer2021comment} it is unlikely that universal recommendations can be made as to which estimators will best balance efficiency and robustness to finite-sample properties in every instance. Instead, estimators should be tested against the analyst’s specific data and estimation goals. To this end, we developed the Realistic Evaluations of Finite sample INference using Efficient Estimators (REFINE2) application which enables analysts to reproduce our evaluation pipeline on their own data: namely, easily input their data, generate plasmode simulations of known structure and effect size, and compare the performance of different estimators and algorithm libraries in line with the simulations we presented in this paper. The  user-friendly Shiny application can be downloaded free at (https://github.com/mengeks/drml-plasmode, \cite{CodeRepo}) and run offline. All that is required is freely-available R software and the Shiny package. An example dataset and models are included, but users can and should load their own datasets.
\begin{figure}[ht]
    \centering
    \includegraphics[width=0.99\linewidth]{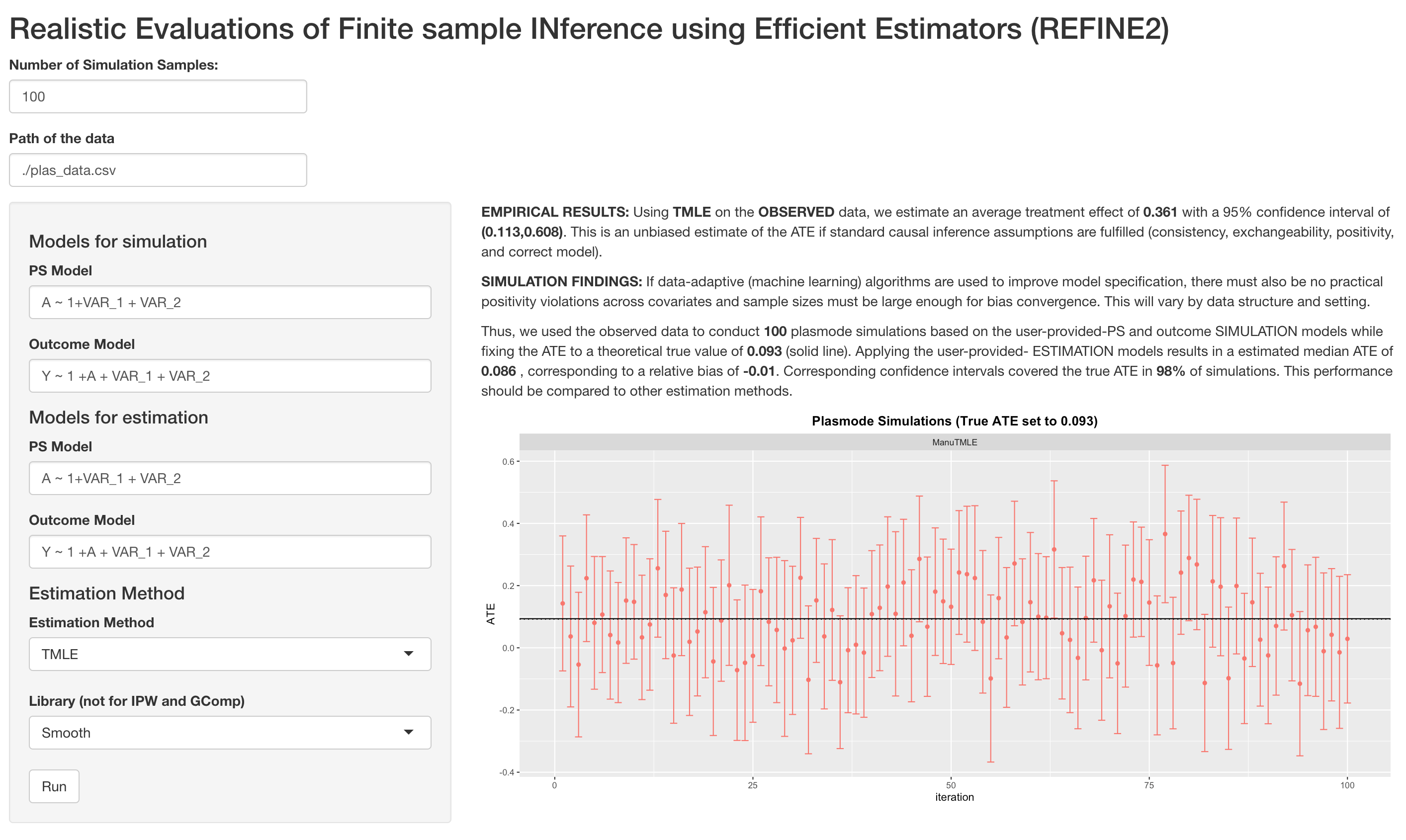}
    \caption{REFINE2: Realistic Evaluations of Finite sample INference using Efficient Estimators}
    \label{fig:REFINE2}
\end{figure}  
\\
\emph{What does it do?}
The REFINE2 tool allows the analyst to check whether the chosen effect estimation strategy, with or without the use of machine learning algorithms, will reliably estimate an unbiased average treatment effect (ATE) and appropriate confidence intervals given their specific dataset and user-provided parameters. Namely, the user provides: 
\begin{itemize}
    \item the dataset  
    \item number of simulation draws
    \item \textbf{Simulation} models representing the "true" relationships between treatment, outcome, and covariates (in R syntax)
    \item \textbf{Estimation} models representing how the effect would be estimated
    \item chosen estimator (\emph{e.g.} IPW, DC-TMLE)
    \item library used to fit the estimator (smooth, non-smooth)
\end{itemize}

Allowing the user to specify both Simulation and Estimator models ensures maximum flexibility on degree of misspecification. For example, variables could be used to simulate the truth but be omitted or mismeasured in the estimation model to test sensitivity. Using the exact same Simulation and Estimation models will test the finite sample / bias convergence properties of the algorithms on the dataset at hand.\\

\emph{Interpreting output}
The tool provides several useful outputs. First, without conducting any simulations, REFINE2 provides a fit of the Estimation model using the specified estimator and alogorithms on the observed data to computes the empirical ATE assuming the required causal identification and estimation assumptions are fulfilled. This is useful for analysts who just want to deploy the algorithms. 
\\
Second, REFINE2 simulates the specified number of datasets using the Simulation models provided while fixing the true ATE. To reduce the similarity between simulation and estimation models noted above, we use a naive estimator (Y ~ X) to choose a reasonable true ATE. Next, REFINE2 fits the Estimation models using the chosen estimator and algorithms on each of the simulated steps and reports summary performance measures. This can be then be repeated and compared across different combinations of estimators and algorithms. The simulated data will be stable for any given dataset and set of Simulation models. \\  

\emph{Future versions and updates}
The current version of REFINE2 supports estimation of ATEs using the estimators and learning libraries implemented in this study. Future updates will include additional estimators, customizable SuperLearner libraries, additional estimands (e.g. risk / rate ratios), data generating approaches, and imputation approaches for missing data.

\subsection{Conclusions}

Our results reinforce the growing call for more thorough evaluation of estimators (\cite{boulesteix2017}), particularly in settings close to those where they will be deployed \cite{balzer2021comment}. Uncommonly considered characteristics such as preservation of observed total treatment and outcome variation from the source data \cite{stokes2020} can be better assured when only simulating parts of the data generating process and drawing the rest of the covariance matrix as observed. As suggested by \cite{bahamyirou2019understanding}, these simulation approaches can be used more routinely to understand the performance of the estimator for the given analytic context. From our initial establishing simulation, we suggest that past evaluations of these estimators, while demonstrating the potential benefit of doubly-crossfit efficient estimators, did not present sufficiently challenging conditions: differences in performance were less than 10 percent (for both bias and variance), consistent with past studies. However, when we evaluate estimators on sets simulated from real data, drops in performance were much more dramatic, particularly with respect to confidence interval coverage. In such settings, we find across numerous scenarios that crossfit efficient estimators fit with smooth models tend to be the optimal compromise -- in the settings where more flexible estimators show a mild benefit, they also have conservative variance estimates at the additional cost of excessive computation time. In settings where numerous effect estimates are desired, such as in bioinformatics, molecular epidemiology, and clinical real-world evidence studies this may be prohibitive. Moreover, we found that even when joining smooth with non-smooth estimators, performance tended to be worse then when only smooth learners were used, a result hinted by \cite{balzer2021comment}. Consequently, as \cite{balzer2021comment} we recommend both routine adoption of real-data-based simulation to evaluate estimator performance, and potentially blinded simulation as recommended by \cite{boulesteix2017}, as well as first considering simpler, stable, smooth models for nuisance function in crossfit estimators for typical clinical studies.

\section*{Acknowledgements}
	
	This research was supported in part by MOH-000550-00 (MOH-OFYIRG19nov-0008) from the Singapore National Medical Research Council to JYH and by U.S. National Institutes of Health grants R01AA23187 and P41EB028242 
	The authors thank Susan Murphy for kind support and Paul Zivich for helpful guidance and comments. The content is solely the responsibility of the authors and does not necessarily represent the official views of the funding agencies.

\bibliography{references}  %%% Remove comment to use the external .bib file (using bibtex).

\section*{Appendix: Exploratory Analysis of the Dataset}
\begin{figure}[h!]
    \centering
    \includegraphics[width=0.7\linewidth]{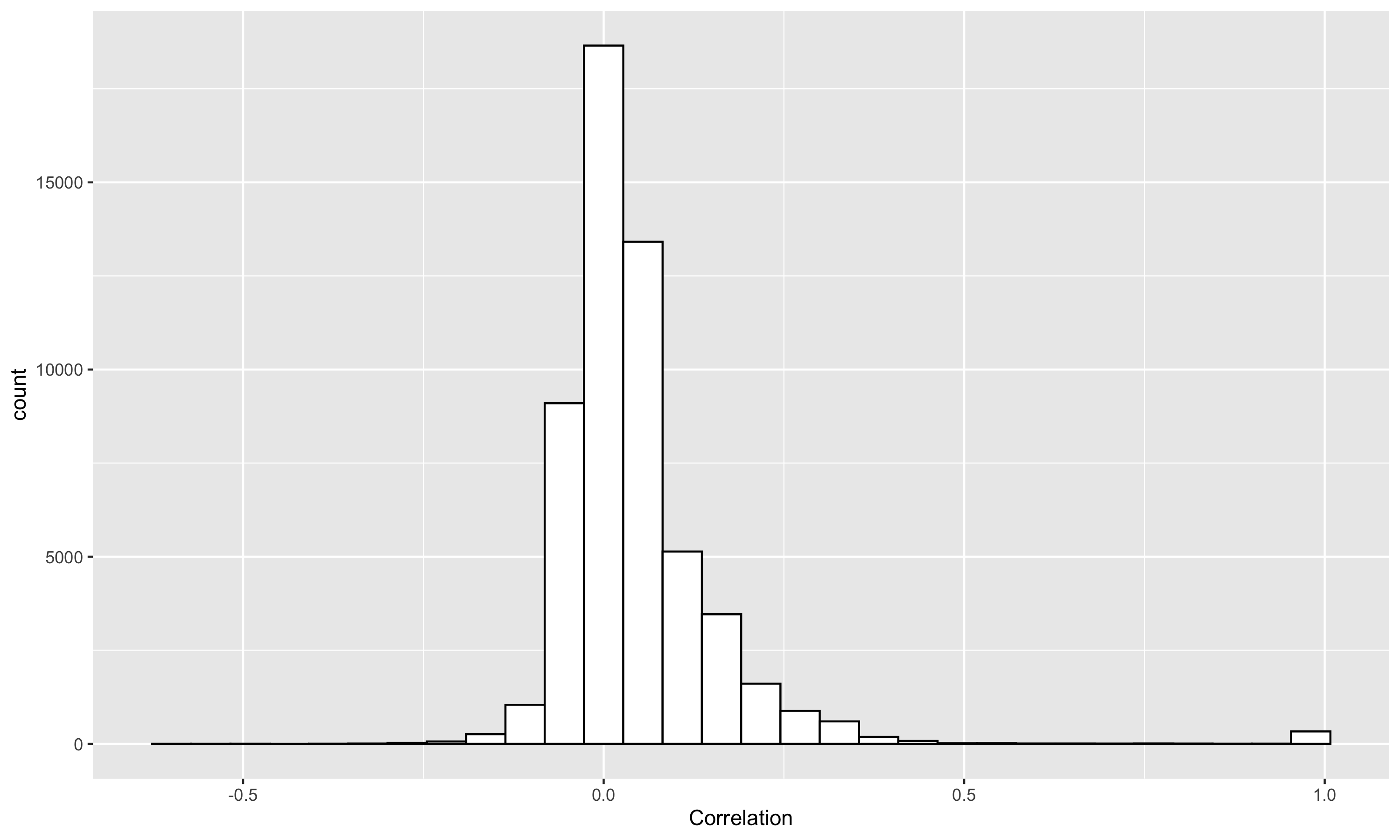}
    \caption{Histogram for Pairwise Correlations}
    \label{fig:corr_hist}
\end{figure}

We present the histogram  of  $\binom{331}{2} = 54615$ pairs of correlation in Figure \ref{fig:corr_hist}. Among them, $30$ of these pairs have absolute value greater than $0.7$, $13$ pairs greater than $0.8$, $4$ pairs greater than $0.9$. 

\section*{Appendix: Result When Different Number of Double Crossfit Procedures are Used}
 Increasing the number of double-crossfit procedures (num\_cf) (Table \ref{tab:A.cfnum}) from 5 to 10 or 20 did not substantially change the relative performance of any of the estimators, though of course increasing run time proportionally. Hence we decided to proceed with num\_cf=5 for all scenarios.
\begin{table}[h!]
		\begin{center}
			\caption{Comparison of performance and run time by number of double-crossfitting procedures. }
			\bigskip
			\label{tab:A.cfnum}
			\begin{tabular}{rcccccc}
				\toprule
			& \multicolumn{2}{c}{DC-TMLE} &   \multicolumn{2}{c}	{DC-AIPW }\\
					\cmidrule(r){2-3} \cmidrule(l){4-5}  
				&   \multicolumn{1}{c}{S} & \multicolumn{1}{c}{NS} & \multicolumn{1}{c}{S} & \multicolumn{1}{c}{NS} \\
				\midrule
\multicolumn{1}{l}{num\_cf=5}	 & & & \\
Bias ($\times$ 100) & 10.96 & 28.49 & 1.91 & -14.71\\  [2pt] 
 SE  & 0.17 & 0.16 & 0.15 & 0.12\\  [2pt] 
 CI covg.  & 0.76 & 0.55 & 0.89 & 0.73\\  [2pt] 
 BVar  & 0.06 & 0.09 & 0.03 & 0.03\\  [2pt] 
 Time & 120.09 & 203.85 & 119.65 & 210.38\\  [10pt] 
\multicolumn{1}{l}{num\_cf=10}	 & & & \\
Bias ($\times$ 100) & 11.1 & 21.99 & 0.23 & -15.08\\  [2pt] 
 SE  & 0.17 & 0.19 & 0.15 & 0.12\\  [2pt] 
 CI covg.  & 0.83 & 0.7 & 0.92 & 0.76\\  [2pt] 
 BVar  & 0.05 & 0.06 & 0.03 & 0.03\\  [2pt] 
Time  & 239.58 & 405.42 & 237.02 & 419.68\\  [10pt] 
 \multicolumn{1}{l}{num\_cf=20}	 & & & \\
 Bias ($\times$ 100) & 7.83 & 25.85 & -0.03 & -15.69\\  [2pt] 
 SE  & 0.18 & 0.19 & 0.15 & 0.13\\  [2pt] 
 CI covg.  & 0.81 & 0.66 & 0.91 & 0.72\\  [2pt] 
 BVar  & 0.05 & 0.06 & 0.03 & 0.03\\  [2pt] 
Time  & 477.01 & 810.81 & 471.6 & 838.44\\  [2pt]
				\bottomrule 
			\end{tabular}
		\end{center}
	\end{table}

\section*{Appendix: True Coefficients Used in Plasmode}
 In generating outcome (OM), there is a term specifying coefficient of exposure variable to 6.6. We present the rest coefficients in the table below. VAR\_i corresponds to $X_i$ for all i=1,2,...

\begin{table}[ht]
\centering
\caption{Scenario A: A.cor uses all variables in OM and PS; A.no.int does not have the interaction terms, i.e., lower 10 terms in OM and PS; A.less.1st only includes $ (VAR\_1 + VAR\_2 + VAR\_5 + VAR\_18 + VAR\_217 )^2$ and VAR\_i, $i\in \{4,7,9,16,17,27,28,31,34,39\}$}
\begin{tabular}{rrr}
  \hline
 & OM Coef & PS Coef \\ 
  \hline
(Intercept) & -3.66 & -0.29 \\ 
  VAR\_1 & 0.83 & -0.61 \\ 
  VAR\_2 & 0.03 & 0.24 \\ 
  VAR\_5 & 0.10 & 0.06 \\ 
  VAR\_18 & 0.15 & 0.11 \\ 
  VAR\_217 & 0.29 & -1.63 \\ 
  VAR\_3 & -0.03 & -0.09 \\ 
  VAR\_4 & -0.00 & 0.04 \\ 
  VAR\_6 & -0.12 & -0.42 \\ 
  VAR\_7 & 0.01 & 0.17 \\ 
  VAR\_8 & 0.02 & -0.06 \\ 
  VAR\_9 & 0.38 & 0.07 \\ 
  VAR\_10 & 0.02 & 0.22 \\ 
  VAR\_11 & -0.01 & -0.16 \\ 
  VAR\_12 & 0.00 & 0.21 \\ 
  VAR\_13 & 0.07 & 0.20 \\ 
  VAR\_14 & 0.03 & 0.16 \\ 
  VAR\_15 & -0.00 & -0.35 \\ 
  VAR\_16 & 0.34 & 0.07 \\ 
  VAR\_17 & 0.05 & 0.17 \\ 
  VAR\_19 & -0.01 & 0.30 \\ 
  VAR\_20 & 0.01 & -0.07 \\ 
  VAR\_21 & 0.00 & 0.03 \\ 
  VAR\_22 & 0.01 & 0.09 \\ 
  VAR\_23 & -0.00 & -0.08 \\ 
  VAR\_24 & -0.00 & -0.16 \\ 
  VAR\_25 & 0.02 & 0.15 \\ 
  VAR\_26 & -0.01 & -0.00 \\ 
  VAR\_27 & 0.00 & 0.00 \\ 
  VAR\_28 & -0.01 & -0.01 \\ 
  VAR\_29 & -0.00 & 0.01 \\ 
  VAR\_30 & -0.03 & -0.59 \\ 
  VAR\_31 & 0.00 & 0.00 \\ 
  VAR\_32 & 0.06 & 0.27 \\ 
  VAR\_33 & 0.04 & -0.62 \\ 
  VAR\_34 & 0.00 & -0.01 \\ 
  VAR\_35 & 0.01 & 0.16 \\ 
  VAR\_36 & 0.00 & -0.00 \\ 
  VAR\_37 & 0.06 & 0.08 \\ 
  VAR\_38 & -0.00 & 0.00 \\ 
  VAR\_39 & -0.09 & 0.11 \\ 
  VAR\_40 & 0.52 & -0.53 \\ 
  VAR\_1:VAR\_2 & -0.01 & 0.01 \\ 
  VAR\_1:VAR\_5 & -0.18 & 0.08 \\ 
  VAR\_1:VAR\_18 & -0.17 & -0.13 \\ 
  VAR\_1:VAR\_217 & 0.03 & 0.64 \\ 
  VAR\_2:VAR\_5 & 0.04 & 0.14 \\ 
  VAR\_2:VAR\_18 & -0.02 & -0.19 \\ 
  VAR\_2:VAR\_217 & 0.03 & -0.00 \\ 
  VAR\_5:VAR\_18 & 0.04 & 0.18 \\ 
  VAR\_5:VAR\_217 & -0.06 & 0.35 \\ 
  VAR\_18:VAR\_217 & 0.00 & -0.02 \\ 
   \hline
\end{tabular}

\end{table}

\begin{figure}
    \centering
    \includegraphics[width=0.9\linewidth]{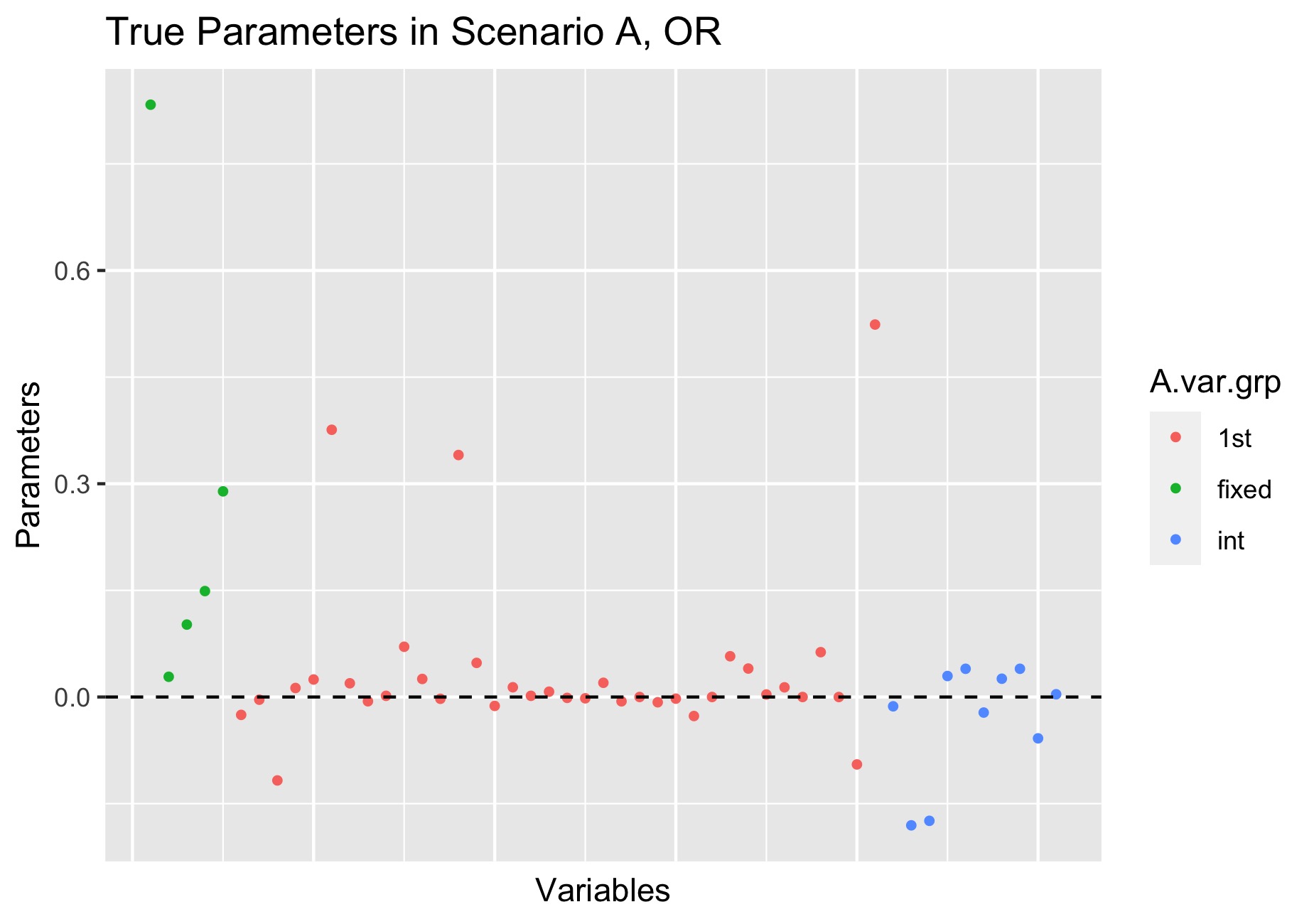}
    \caption{Plot for Parameters in Scenario. Fixed means those in all 3 simulations, i.e., first order terms for variable 1, 2,5,18,217;1st means all other first order terms in variables 1 to 40; int means interaction terms among variables 1, 2,5,18,217 }
    \label{fig:a-param}
\end{figure}

\begin{table}[ht]
\centering
\caption{Scenario B coefficients}
\begin{tabular}{rrr}
  \hline
 & OM Coef & PS Coef \\ 
  \hline
(Intercept) & -3.77 & -10.35 \\ 
  VAR\_1 & 0.89 & 0.22 \\ 
  VAR\_2 & 0.13 & 0.67 \\ 
  VAR\_5 & 0.06 & 0.45 \\ 
  VAR\_18 & 0.21 & 0.62 \\ 
  VAR\_217 & 0.29 & -1.77 \\ 
  VAR\_34 & 0.00 & -0.00 \\ 
  VAR\_27 & -0.00 & -0.00 \\ 
  VAR\_4 & -0.01 & -0.02 \\ 
  VAR\_31 & -0.00 & -0.00 \\ 
  VAR\_28 & -0.01 & 0.00 \\ 
  VAR\_17 & 0.05 & 0.21 \\ 
  VAR\_16 & 0.34 & 0.09 \\ 
  VAR\_9 & 0.50 & 0.68 \\ 
  VAR\_7 & 0.05 & 0.04 \\ 
  VAR\_39 & -0.09 & -0.01 \\ 
  VAR\_1:VAR\_2 & -0.03 & -0.10 \\ 
  VAR\_1:VAR\_5 & -0.19 & -0.10 \\ 
  VAR\_1:VAR\_18 & -0.17 & -0.03 \\ 
  VAR\_1:VAR\_217 & 0.04 & 0.45 \\ 
  VAR\_2:VAR\_5 & 0.04 & 0.11 \\ 
  VAR\_2:VAR\_18 & -0.01 & 0.02 \\ 
  VAR\_2:VAR\_217 & 0.01 & 0.12 \\ 
  VAR\_5:VAR\_18 & 0.02 & -0.13 \\ 
  VAR\_5:VAR\_217 & -0.05 & 0.43 \\ 
  VAR\_18:VAR\_217 & 0.01 & 0.02 \\ 
   \hline
\end{tabular}
\end{table}

\begin{table}[ht]
\centering
\caption{Scenario C Coefficients}
\begin{tabular}{rrr}
  \hline
 & OM Coef & PS Coef \\ 
  \hline
(Intercept) & 12.49 & -4.86 \\ 
  VAR\_1 & 1.16 & 0.38 \\ 
  VAR\_2 & -0.37 & 0.63 \\ 
  VAR\_5 & -0.19 & 0.64 \\ 
  VAR\_18 & -0.01 & 0.45 \\ 
  VAR\_217 & 0.64 & -1.64 \\ 
  A1:VAR\_1 & -1.37 & 0.00 \\ 
  A1:VAR\_2 & 0.41 & 0.00 \\ 
  A1:VAR\_5 & -1.89 & 0.00 \\ 
  A1:VAR\_18 & 0.04 & 0.00 \\ 
  A1:VAR\_217 & 1.24 & 0.00 \\ 
  VAR\_1:VAR\_2 & -0.08 & -0.13 \\ 
  VAR\_1:VAR\_5 & -0.26 & -0.18 \\ 
  VAR\_1:VAR\_18 & -0.19 & -0.03 \\ 
  VAR\_1:VAR\_217 & 0.02 & 0.45 \\ 
  VAR\_2:VAR\_5 & 0.27 & 0.14 \\ 
  VAR\_2:VAR\_18 & -0.00 & 0.03 \\ 
  VAR\_2:VAR\_217 & 0.03 & 0.12 \\ 
  VAR\_5:VAR\_18 & 0.11 & -0.06 \\ 
  VAR\_5:VAR\_217 & -0.17 & 0.37 \\ 
  VAR\_18:VAR\_217 & 0.08 & 0.03 \\ 
   \hline
\end{tabular}
\end{table}
\end{document}